\documentclass[acmtog,  table]{acmart}
\acmSubmissionID{502}

\usepackage{amsmath}
\usepackage{booktabs} 
\usepackage{float}
\usepackage{layouts}
\usepackage{wrapfig}
\usepackage[ruled,vlined]{algorithm2e}

\usepackage{bbm}

\usepackage{appendix}

\definecolor{forestgreen}{rgb}{0.13, 0.55, 0.13}
\definecolor{lava}{rgb}{0.81, 0.06, 0.13}
\definecolor{magenta}{rgb}{0.7, 0.0, 1.0}

\newcommand{\edit}[1]{ #1}

\definecolor{staticColor}{HTML}{005AB5}
\newcommand{\staticp}[1]{\textcolor{staticColor}{#1}}
\def\staticColorName{blue}
\definecolor{dynamicColor}{HTML}{DC3220}
\newcommand{\dynamicp}[1]{\textcolor{dynamicColor}{#1}}
\def\dynamicColorName{red}

\newcommand{\reffig}[1]{Fig.~\ref{fig:#1}}
\newcommand{\refeq}[1]{Eq.~(\ref{eq:#1})}
\newcommand{\refsec}[1]{Sec.~\ref{sec:#1}}

\newcommand*{\img}[1]{%
    \raisebox{-0\baselineskip}{%
        \includegraphics[
        keepaspectratio,
        ]{#1}%
    }%
}

\newcommand*{\timestamp}[2][-0.5cm]{%
   \sffamily \small%
  \vspace{#1}%
  \begin{flushright}%
  \protect\img{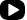}#2%
  \end{flushright}%
}
\def\tsAquarium{0m40s}
\def\tsCoarseningMeshes{3m20s}
\def\tsFullvsReduced{0m16s}
\def\tsMorayModalDerivs{4m57s}

\def\tsBingbyLocalRotation{3m34s}
\def\tsSubspaceComparison{1m45s}
\def\tsCthuluLocalMinimum{1m27s}
\def\tsBarFieldVis{1m21s}
\def\tsWeightVis{1m56s}
\def\tsAffineMotions{2m08s}
\def\tsHeterogeneousMaterialExperiment{4m01s}
\def\tsBarModalWarpingComparison{3m52s}
\def\tsGiraffeRSComparison{3m44s}
\def\tsConstrainedVsUnconstrained{4m27s}
\def\tsCrocodileCorotVsARAP{4m38s}
\def\tsRigidBodyGame{5m42s}
\def\tsMixamoRigSwitchingApp{5m04s}

\def\tsMediaPipeFace{6m02s}
\def\tsMediaPipePose{6m11s}
\usepackage{amsmath}

\DeclareMathOperator*{\argmin}{argmin}
\usepackage{mfirstuc}
\MFUnocap{et}
\MFUnocap{al}

\newcommand{\A}{\boldsymbol{A}}
\newcommand{\Bdisp}{\boldsymbol{B}_\text{disp}}
\newcommand{\Blbs}{\boldsymbol{B}_\text{lbs}}
\newcommand{\B}{\boldsymbol{B}}
\newcommand{\I}{\boldsymbol{I}}
\newcommand{\J}{\boldsymbol{J}}
\newcommand{\M}{\boldsymbol{M}}
\newcommand{\R}{\boldsymbol{R}}
\newcommand{\T}{\boldsymbol{T}}
\newcommand{\W}{\boldsymbol{W}}
\newcommand{\X}{\boldsymbol{X}}
\newcommand{\Y}{\boldsymbol{Y}}
\newcommand{\Z}{\boldsymbol{Z}}
\newcommand{\g}{\boldsymbol{g}}
\newcommand{\p}{\boldsymbol{p}}

\newcommand{\dimp}{p}
\newcommand{\nummodes}{m}

\newcommand{\numtets}{t}
\newcommand{\numclusters}{r}
\newcommand{\x}{\boldsymbol{x}}
\newcommand{\z}{\boldsymbol{z}}

\providecommand{\C}{}
\renewcommand{\C}{\boldsymbol{C}}
\providecommand{\H}{}
\renewcommand{\H}{\boldsymbol{H}}
\providecommand{\M}{}
\renewcommand{\M}{\boldsymbol{M}}
\providecommand{\P}{}
\renewcommand{\P}{\boldsymbol{P}}
\providecommand{\u}{}
\renewcommand{\u}{\boldsymbol{u}}

\newcommand{\rep}[1]{\left(#1 \otimes \boldsymbol{I}\right)}
\newcommand{\repR}{\rep{\R}}

\acmJournal{TOG}

\begin{document}
\title{Fast Complementary Dynamics via Skinning Eigenmodes}

\author{Otman Benchekroun}
\email{otman.benchekroun@mail.utoronto.ca}
\affiliation{
\institution{University of Toronto}
\country{Canada}
}
\author{Jiayi Eris Zhang}
\email{eriszhang@stanford.edu}
\affiliation{
 \institution{Stanford University}
  \country{U.S.A.}
}
\author{Siddhartha Chaudhuri}

\email{sidch@adobe.com}
\affiliation{
 \institution{Adobe Research}
  \country{India}
}
\author{Eitan Grinspun}
  \email{eitan@grinspun.com}
\affiliation{
  \institution{University of Toronto}
  \country{Canada}
}
 \author{Yi Zhou}
 \email{yizho@adobe.com}
\affiliation{
\institution{Adobe Research}
\country{U.S.A.}
}
\author{Alec Jacobson}
 \email{alecjacobson@gmail.com}
\affiliation{
 \institution{University of Toronto and Adobe Research}
 \country{Canada}
}

\begin{abstract}
    We propose a reduced-space elastodynamic solver that is well suited for augmenting rigged character animations with secondary motion. At the core of our method is a novel deformation subspace based on Linear Blend Skinning that overcomes many of the shortcomings prior subspace methods face. Our skinning subspace is parameterized entirely by a set of scalar weights, which we can obtain through a small, material-aware and rig-sensitive generalized eigenvalue problem. The resulting subspace can easily capture rotational motion and guarantees that the resulting simulation is rotation equivariant.  We further propose a simple local-global solver for linear co-rotational elasticity and propose a clustering method to aggregate per-tetrahedra non-linear energetic quantities. The result is a compact simulation that is fully decoupled from the complexity of the mesh. 
\end{abstract}

%
%


%
%


\begin{teaserfigure}
   \centering%
  \includegraphics[width=\textwidth]{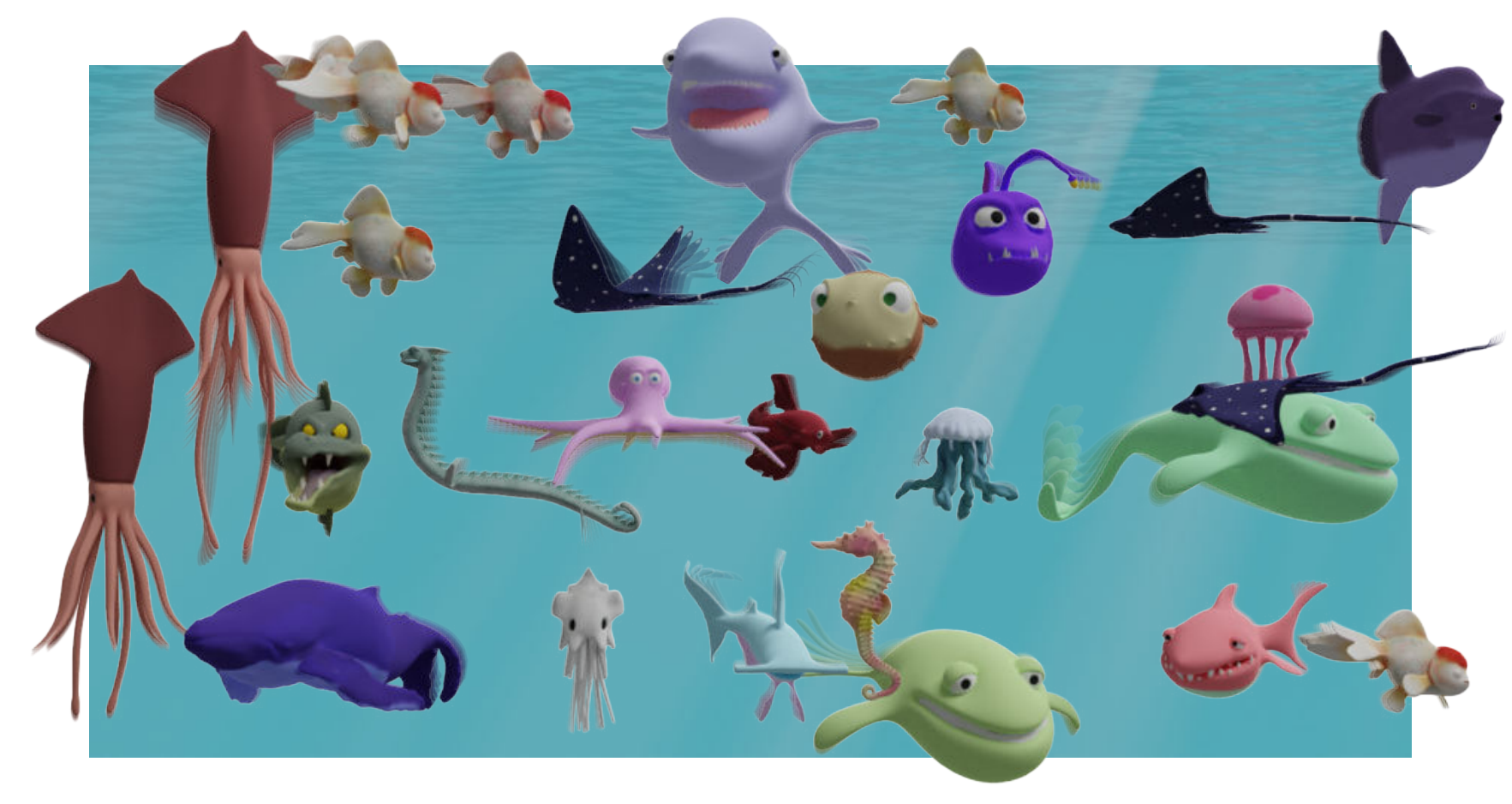}\timestamp{\tsAquarium}%
  \caption{%
  The simple rig motions of 26 underwater sea creatures are augmented with our \emph{real-time} secondary dynamics.
  The full scene of 330,563 mesh vertices and 1,293,625 tetrahedra runs at over 60 fps.
  Throughout our paper, the \protect\img{images/video-symbol.pdf} indicates a corresponding clip in the supplemental video.
  \label{fig:teaser-figure}
  }
\end{teaserfigure}
\maketitle

\section{Introduction}




Virtual reality, video games, and digital art increasingly make use of controllable animated characters. Such characters should provide \emph{interactive} responses to user input in order to communicate the action and emotion of the moment. On the other hand, their motion must be rich with \emph{realistic} visual details, which is what brings these characters to life. 


The most widely used construct for authoring animations are deformation \emph{rigs} mapping low-dimensional parameters (e.g., skeleton or cage positions) to static geometric deformations. Managing these rigs can quickly overwhelm an artist. Simple rigs are easy to animate, but their deformations lack detail; complex rigs provide rich, fine-grained detail, but they are daunting to manipulate.

\citet{Zhang:CompDynamics:2020} offer a way out of this dilemma; they employ a physics simulation whose role is to supplement rig motions with secondary dynamics that are orthogonal --- in the algebraic sense --- to the motion of the rig. The resulting \emph{complementary dynamics} are exactly those motions not producible by the rig itself.

\begin{figure}
\centering\includegraphics{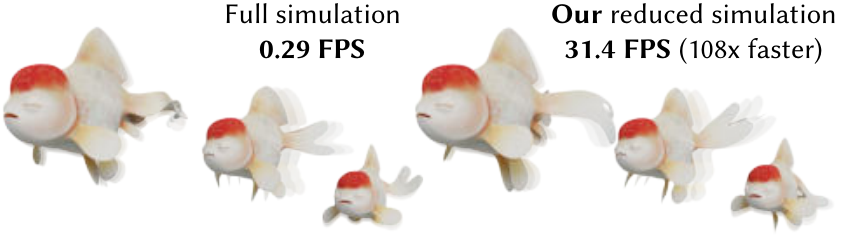}\timestamp[-0.125cm]{\tsFullvsReduced}
    \caption{
\label{fig:full_vs_reduced_cdl} Our reduced complementary dynamics model can reproduce rich visual details at a fraction of the cost of the original method (10,000 vertices, 42,205 tets).}
\end{figure}

\begin{figure}
\includegraphics[width=\linewidth,keepaspectratio]{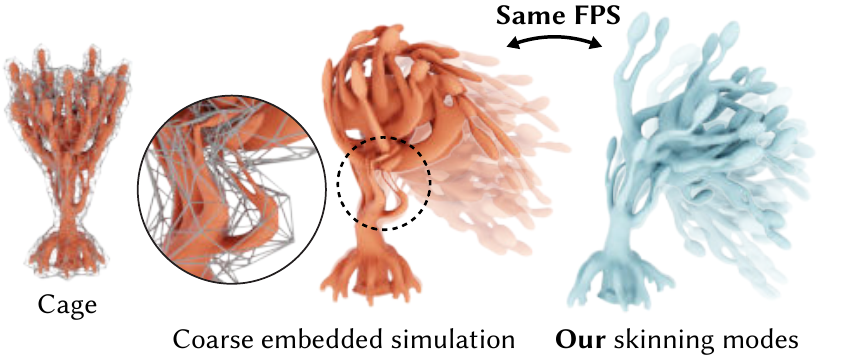}\timestamp[-0.125cm]{\tsCoarseningMeshes}
\caption{A coarse embedded simulation groups motion across vertices that are close in Euclidean space. Instead, our subspace groups motion across vertices that share elastic energy properties (cage constructed via \cite{breaking-good}).
 \label{fig:coarsening-meshes}}
\end{figure}

Unfortunately, complementary dynamics is poorly suited to interactive applications for two reasons. First, enforcing rig complementarity introduces significant computational overhead. Second, the runtime of the method grows with the mesh resolution.  Adding secondary motion even on a modestly sized mesh requires computation that lags far from real-time rates. As shown in \reffig{full_vs_reduced_cdl}, a complementary dynamics animation on a modestly sized mesh of 10,000 vertices, 42,000 tetrahedra requires 3 seconds of compute time for each frame. Attempting to accelerate this method by embedding a high resolution display mesh inside of a coarse simulation (see \reffig{coarsening-meshes}) 
comes at the cost of visible cage artifacts.

To accelerate any high-dimensional optimization problem, a popular approach is to solve the problem in a low-dimensional, representative \emph{subspace}. The \emph{de-facto} subspace for elastodynamics in graphics is the one spanned by the first few eigenvectors of the elastic energy Hessian \cite{PentlandWilliams1989}. We call these eigenvectors \emph{\textbf{displacement modes}} because---for elasticity---they correspond to the set of least energy-incurring infinitesimal displacements about the rest state. 

\begin{figure}[h]
\includegraphics[width=\linewidth]{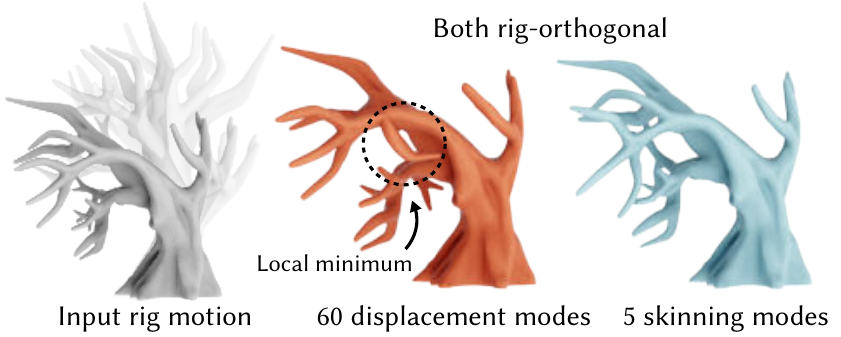}
\caption{Using a vanilla rig-complementary displacement subspace (red) can lead to kinks and local minima when a user rotates the rig. Our rig-complementary skinning eigenmodes (cyan) are rotation equivariant and are much better suited for accomodating this type of motion.
 \label{fig:dead-tree-local-rotation-equivariance}} 
\end{figure}

Our use case exposes the pitfalls of this classical subspace. First, it is well known that this subspace does not represent rotational deformations, leading to warping or shearing artifacts \cite{ModalWarping}. Second and (as we will show) \textit{distinctly}, this subspace induces simulations lacking \emph{rotation equivariance}. 
In a nutshell, a rotation equivariant optimization produces a rotated version of the same minimizer when the problem geometry is rotated. The absence of this important property leads to frame-dependent artifacts. This is particularly noticeable in our application, where local rotations form a primary degree of freedom of the interactive rig. As the user interacts with the rig, the secondary physics exhibit non-physical dampened motion, and the mesh gets stuck in local minima as shown in \reffig{dead-tree-local-rotation-equivariance} and \reffig{random-init-linear-subspace}.

To address these challenges, we propose \textbf{\textit{skinning eigenmodes}} for reduced simulation. Inspired by linear blend skinning, the subspace spanned by our skinning modes yields rotation equivariant secondary elastodynamics. Indeed, we \textit{prove} that it meets the necessary and sufficient conditions for doing so. Our subspace is fully parameterized by a compact set of \emph{skinning weights}, which we derive through a physically motivated, material aware and simple to implement generalized eigenvalue problem. 

The formulation of our skinning modes as the solution to an eigenvalue problem has many advantages. A user can easily explore the cost versus richness tradeoff of the resulting dynamics by simply truncating the eigenspace. We also benefit from a large array of work that aims at promoting different qualities from eigen problems, such as enforcing locality and sparsity in our modes \cite{Brandt2017CompressedVibrationModesofElasticBodies, Nasikun2018, compressedModesADMM}, or enforcing homogeneous linear equality constraints \cite{golub1973}. 
In particular we benefit from the latter to make our skinning subspace orthogonal to the input rig-space. This effectively makes our modes rig-sensitive, allowing them to more efficiently capture the space of secondary motions available given an input rig.

However, skinning modes alone are insufficient for achieving realistic-looking real-time dynamics. This is because secondary elastodynamics look best with non-linear elastic models. Simulating such materials, even with a deformation subspace,  \emph{still} requires the computation of per-tetrahedron quantities \cite{ModalWarping}. This once again ties our runtime complexity to the resolution of the mesh. To accommodate this, we adopt clustering \cite{Jacobson-12-FAST}, which approximates these per-tet quantities to per-cluster ones. We also extend Jacobson et al.\ \shortcite{Jacobson-12-FAST}'s local-global solver to approximate a co-rotational elastic potential.
The result is an iterative solver that harmonizes well with our subspace and allows rich non-linear secondary motions in a simulation step that is \emph{entirely} decoupled from the mesh resolution. Additionally, by virtue of being based on \emph{linear} blend skinning, projecting from our low dimensional subspace to the full space can be done efficiently in the vertex shader.

\reffig{teaser-figure} demonstrates our reduced elastodynamics augmenting a large scene with secondary motion in \emph{real}-time. We further demonstrate the success of our method through a variety of comparisons to the original (offline) complementary dynamics and against alternative real-time acceleration methods.
We additionally highlight how our skinning modes can be used for a wide range of standard deformation tasks (not just complementary dynamics).
%
Finally, we show successful application of our real-time method to scenarios with rapidly changing rig-input, such as rigid-body enrichment, VR puppetry, and rigged character secondary effects.

\section{Related Work}
There are many choices for accelerating (full-space) elastodynamic simulation, such as Position Based Dynamics \cite{MullerPBD2007}, Projective Dynamics \cite{Bouaziz:2014:Projective}, or Mixed FEM \cite{MixedFEMTrusty2022}. Elastodynamic simulations usually require the solution to a linear system of equations. A simple method of accelerating them is to just use a faster solver such as Multigrid methods \cite{LiuMultigrid2021}, Newton and Quasi-Newton methods \cite{QuasiNewtonLiu2017} and matrix pre-factorizations such as Cholesky decomposition.
Yet another option is to reformulate the constitutive equations to allow for a efficient boundary-only discretization
\cite{artDefoJames1999, Sugimoto:2022:BEM}.
All these methods scale in complexity with the final mesh resolution and scale \emph{at best} linearly with the complexity of the mesh. This puts an unnecessary burden on the user, forcing them to choose between fast elastodynamics with coarse meshes, or slow elastodynamics with fine meshes.
To obtain sub-linear rates the simulation needs to be carried out in a low-dimensional representative subspace.

\subsection{Subspace Simulation}

The most popular subspace for most graphics tasks is the one composed of the first few eigenvectors of the energy Hessian \cite{PentlandWilliams1989}, which we call displacement modes. 

Unfortunately, displacement modes are not well suited for large displacements, and in particular they struggle to capture rotational motion \cite{DyRT, BarbicJames:RealTimeSTVK} . Much work over the last two decades aims at remedying this short-coming for subspace simulation, a goal we share. On top of this however, we also expose another desired property that displacement modes do not generally satisfy: they lead to simulations that are not \textit{rotation equivariant}. 

 \citet{BarbicJames:RealTimeSTVK} introduce modal derivatives, later extended to geometric modelling \cite{Hildebrandt2011} and structural analysis \cite{nonlinearCompliantDuenser2022}. This methodology provides extra derivative modes, whose goal is to correct the primary subspace as it falls out of date with large deformations. While these methods provide vast improvements over traditional displacement modes, the added derivative modes do not represent rotational motion, nor do they ensure rotation equivariance in the resulting simulation.

 To accommodate rotational motion, modes can be warped with best fit aggregate rotational motion \cite{ModalWarping} or skinning motion \cite{EigenSkinKry2022}.
In a similar vein, sub-structuring (also called domain decomposition) \cite{Barbic:2011:RealTimeLargeDefoSubstructuring, KimJamesMultiDomainStitching}, separates a shape into independent regions, each with their own local linear subspace. The rotational motion is tracked externally and is used to update the subspace for each region. 
Instead of updating the quality of the linear basis, Rotation Strain coordinates \cite{RScoords} attempt to \emph{fix} the rotation-lacking motion of the subspace simulation at the end of each time-step via a non-linear projection step. Unfortunately all these methods require per timestep "fixes", which result in a costly simulation step that limits the richness of the dynamics available for real-time interaction.

The data driven neural subspaces of \citet{zheng2021deep} show promise for real-time applications, but suffer from artifacts when applied on meshes they are not trained on. They also provide no guarantee of any energy conserving properties desired of an elasticity subspace.

Von Tycowicz \shortcite{Tycowicz2013} expand traditional displacement modes with each of the $d \times d$ entries of a linear map. Their result also leads to a rotation spanning subspace that guarantees simulation rotation equivariance. However, their subspace cannot be trivially made to accommodate homogeneous linear equality constraints, such as the one needed to impose rig-complementarity. As a result, this subspace is inefficient for real-time complementary dynamics. 

\subsection{Subspace Simulation via Skinning Modes}
Linear blend skinning is a popular rigging method used to easily pose characters. \edit{ It has also widely been used as a subspace for deformation,  with prior works varying in how they compute their skinning weights. While prior methods acknowledge that skinning subspaces produce high quality rotational deformation, we further 
 motivate our use of this subspace by identifying that it \emph{guarantees} rotation equivariance, a key invariant for elastica that pure displacement subspaces break. 
} 

\edit{\citet{rohmer2021velocity} propose a method that prescribes secondary motion directly on the user-provided primary rig. They derive physics-like behavior that aims to mimic common squash and stretch motions. 
For more physically motivated inverse kinematics, 
\citet{Jacobson-12-FAST} use the user provided skinning rig as a subspace for minimizing an elasto-static energy. Rig-Space Physics \cite{RSPHahn} aim to add secondary motion in a rig subspace, a similar goal to ours. These works all rely on an artist to specify the rig subspace for secondary motion themselves. \citet{hahn2013efficient} mitigates this by fitting the rig subspace to a target simulation using a least-squares solve. However, this still requires a previously constructed rig animation as input.  In contrast, we propose a method that derives skinning weights entirely from the rest-pose geometry of our shape.
}

One way to achieve such weights is by requiring them to be smooth. To this end,
\citet{1Gilles2011, Wang:2015:LinearSubspaceDesign} derive skinning weights as a solution to a Laplace and bi-Laplace equation respectively. \edit{Similarly,  \citet{Lan2020, Lan2021} use bounded bi-harmonic weights \cite{JacobsonBPS11} to create smooth, non-negative skinning weights. Instead, \citet{Brandt2018HyperReducedPD, brandt2019}
 compute weights with simple truncated radial basis functions to obtain similarly smooth and locally supported bases.} While smoothness is an attractive quality for deformation, it is ill suited to represent motions for more complex materials with large heterogeneities. 
\citet{Faure2011} address this by sampling source points for their weight computation via a \emph{compliance} (inverse stiffness) weighed sampling of their shape, for which they rely on a Voronoi tessellation of their shape. \edit{Additionally, just like \citet{Tycowicz2013}, it is unclear how to enforce the complementarity constraint on the construction of these skinning weights.
 In contrast, our weights are derived from a generalized eigenvalue problem on a weight-space Hessian, which allows them to reflect material properties and accommodate the rig-complementarity constraint without requiring additional discretization. }

\begin{figure}[!t]
\includegraphics[width=\linewidth]{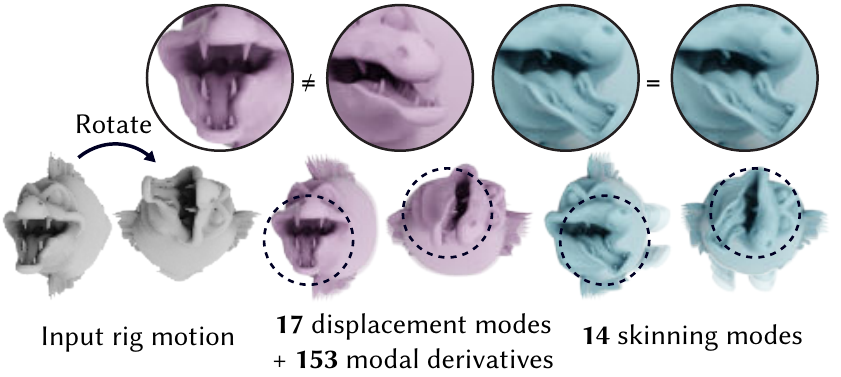}\timestamp[-0.125cm]{\tsMorayModalDerivs}
 \caption{Our skinning eigenmodes' closure under rotations ensures that our subspace complementary dynamics simulations are rotation equivariant. A user experiences the same dynamics independent of the rig orientation. The same cannot be said for a \edit{complementary} subspace built from displacement modes + modal derivatives (see Section \ref{sec:rotating_frames}).
  \label{fig:moray-rotation-equivariance}} 
\end{figure}

\subsection{Simulations Embedded in Rotating Frames}
\label{sec:rotating_frames}
\edit{
A common solution for fixing some of the global rotational artifacts that occur in free-flying elastodynamic simulation, (as shown in \reffig{subspace-comparisons} or \reffig{random-init-linear-subspace}) is to embed the elastodynamic simulation in a rotating frame \cite{Terzopoulos1988}. Here, the elastic response is computed in a rest frame, while the rotation is tracked explicitly via a rigid body simulator \cite{Terzopoulos1988}. The two simulations are then coupled with forces that arise out of angular momentum \cite{DyRT}, and the rotation is then used to transform the deformed rest state every time-step.
Unfortunately, this approach doesn't easily generalize for complementary dynamics since the rigid motions are not orthogonal to the rig, which is why the artifacts present in \reffig{moray-rotation-equivariance} would remain. For rigs that have a global rotation, one could use that as the rotational degree of freedom \cite{DyRT}, unfortunately most rigs do not directly have such a global rotation.}


\section{Background: Complementary Dynamics}
Complementary dynamics provides a methodology for augmenting rigged animations with the detailed  elastodynamics \cite{Zhang:CompDynamics:2020}.
They split the total displacement field $\boldsymbol{u} \in \mathbb{R}^{n(d)}$ for a mesh with $n$ vertices in $d$-dimensional space into an artist prescribed component $\boldsymbol{u}^r$, and a physical component $\boldsymbol{u}^c$:
\begin{align}
    \boldsymbol{u} = \boldsymbol{u}^r + \boldsymbol{u}^c.
\end{align}

The artist-prescribed component is obtained from a rig, which takes as input low dimensional rig parameters $\boldsymbol{p} \in \mathbb{R}^\dimp$
that are exposed to the user for interactive manipulation, and maps them to high dimensional rig displacement $\boldsymbol{u}^r$:
\begin{align}
    \boldsymbol{u}^r = f_{\text{rig}}(\boldsymbol{p}).
\end{align}

The physical motion on the other hand is obtained from a physics simulation, which can be formulated as an energy minimization problem:
 \begin{align}
 \label{eq:cd}
     \boldsymbol{u}^c = \argmin_{\boldsymbol{u}^c} E(\boldsymbol{u}^c + \boldsymbol{u}^r + \boldsymbol{x}_0) \quad \text{s.t.}\ 
    \boldsymbol{J}^T \boldsymbol{u}^c = \boldsymbol{0},
 \end{align}
We introduce the rig Jacobian  $\boldsymbol{J} = \frac{\partial f_{\text{rig}}(\boldsymbol{p}) }{\partial \boldsymbol{p}} \in \mathbb{R}^{n(d)\times \dimp }$.
Note that complementary dynamics is completely agnostic to the elasto-dynamic energy used $E(\cdot)$. Where it differs from a regular elasto-dynamic energy minimization is the specification of the complementarity constraint $\boldsymbol{J}^T \boldsymbol{u}^c = \boldsymbol{0}$. This constraint enforces that the physical motion must not be in the space of motions producible by the rig, $\mathrm{Col}(\boldsymbol{J})$. 
\citet{Zhang:CompDynamics:2020} enforce the complementarity constraint through Lagrange multipliers and the resulting energy is minimized using Newton's method, requiring the frequent solve of the following KKT system:
\begin{align}
    \begin{bmatrix}
    \boldsymbol{H} & \boldsymbol{J} \\
    \boldsymbol{J}^T & \boldsymbol{0}
    \end{bmatrix}
    \begin{bmatrix}
    d\boldsymbol{u}^c \\
    \boldsymbol{\lambda} 
    \end{bmatrix} = 
    \begin{bmatrix}
    -\boldsymbol{g} \\
    \boldsymbol{0} 
    \end{bmatrix},
\end{align}
where $\boldsymbol{H} \in \mathbb{R}^{n(d) \times n(d) }$ 
and
$\boldsymbol{g} \in \mathbb{R}^{n(d)}$ 
are
the elasto-dynamic energy Hessian \& gradient, $d\boldsymbol{u}^c \in \mathbb{R}^{n(d)}$ is the Newton search direction, and $\boldsymbol{\lambda} \in \mathbb{R}^\dimp$ collects the Lagrange multipliers enforcing the complementarity constraint.

Iteratively 
solving this system is too expensive in real-time applications for two main reasons:
\begin{enumerate}
  \item the system scales with mesh resolution, and
  \item the constraints $\boldsymbol{J}^T \boldsymbol{u}^c = \boldsymbol{0}$ are typically dense, even if $\boldsymbol{H}$ is sparse.
\end{enumerate}

\section{Skinning Eigenmodes}
\label{sec:skinning-eigenmodes}

Our goal is to derive a suitable linear subspace so that 
full-space complementary displacements $\u^c \in \mathbb{R}^{n(d)}$ may be approximated with a smaller $m$-dimensional linear subspace:
\begin{align}
\u^c \approx  \B \z,
\label{eq:subspace-approx}
\end{align}
where the columns of $\B \in \mathbb{R}^{n(d) \times m}$ form the subspace basis, and the vector $\z \in \mathbb{R}^{m}$ are the reduced degrees of freedom optimized at run-time.

To apply subspace reduction to the complementary dynamics problem (\refeq{cd}), we would like our subspace $\B$ to simultaneously: deal well with large (global and local) rotations, well-approximate the space of low-energy displacements, and accommodate the rig-complementarity constraints.
\edit{We make use of a linear blend skinning subspace basis for deformation $\Blbs$ \cite{1Gilles2011} and demonstrate in the following sections how we meet these three desirable criteria.}

Linear blend skinning represents displacements as a weighted summation of $m$ affine transformations applied to a shape's rest positions. The $i$th vertex on the shape is displaced via
\begin{align}
\u_i = \sum_{b=1}^m  w_{ib} \T_b \X_i,
\end{align}
where $w_{ib} \in \mathbb{R}$ is the weight of the $b$th transformation at vertex $i$, $\T_b \in \mathbb{R}^{d \times (d+1)}$ is the $b$th transformation, and $\X_i \in \mathbb{R}^{d+1}$ is the $i$th vertex's rest position in homogeneous coordinates.
This equation is linear in $\T$ and so it follows that it may be rearranged so that the degrees of freedom in $\T$ are collected in a single vector $\z = \text{vec}(\T) \in \mathbb{R}^k$ with $k=d(d+1)m$ and the 
weights $w$ and rest positions $\X$ form the columns of a matrix $\Blbs \in \mathbb{R}^{n(d) \times k}$:
\begin{align}
\Blbs &= \I_{d} \otimes (( \boldsymbol{1}_m^T \otimes \X ) \odot ( \W \otimes \boldsymbol{1}_{d+1}^T) ) \ ,
    \label{eq:linear-blend-skinning-matmul}
\end{align}
where $\W \in \mathbb{R}^{n \times m}$ is a matrix with columns collecting each transformation's weights and 
$\X \in \mathbb{R}^{n \times (d+1)}$ collects homogeneous rest positions in rows.
Since the rest positions are generally given, the only variables in our subspace design are the weights $\W$. We now propose a method for choosing $\W$ to ensure that weights span low-energy motions \emph{and} satisfy the complementarity constraints by construction. We defer discussion of how our choice of linear blend skinning subspace directly ensures good rotational properties (see \refsec{rotations}).

Our first step follows the process for standard modal subspaces.
We approximate our elastodynamic energy with a Taylor expansion about the rest state truncated to second order terms,
\begin{align}
    E(\u + \x_0) &=  E_0 + \u^T \g_0 + \frac{1}{2} \u^T \H_0 \u + \mathcal{O}(\left\|\u\right\|^3) \, ,\nonumber \\
    E(\u + \x_0) &\approx  \frac{1}{2} \u^T \H \u \, ,
\end{align}
where we have dropped the subscript for the elastic energy Hessian in the second line for readability.
Without loss of generality, we have assumed zero elastic energy and vanishing elastic forces at rest ($E_0 = 0, \, \g_0 = \boldsymbol{0}$). 

We arrive at a standard modal subspace by
adding a (mass-) orthogonality constraint  $\B^T \M \B = \I$; substituting the subspace $\u = \B \z$; assuming $\z \sim \mathcal{D}$ are sampled from an as-of-yet arbitrary distribution, and minimize the expected value of the energy over $\B$:
\edit{
\begin{align}
   \Bdisp =  &\argmin_{\B^T \M \B = \I}  \mathbb{E}_ {\z\sim \mathcal{D}} [\z^T \B^T  \H \B \z] \\ 
    =  & \argmin_{\B^T \M \B = \I} \mathrm{tr}(\B^T \H \B \mathbb{E}_ {\z \sim \mathcal{D} }[\boldsymbol{zz}^T]) \  . 
\end{align}
}
We further assume that $\z$ are independent and identically distributed (i.i.d.) samples of a normal distribution, then  $\mathbb{E}_ {\z  \overset{\text{i.i.d.}}{\sim} \mathcal{N}( \boldsymbol{0}, \boldsymbol{1})} [\boldsymbol{zz}^T] = \I$, and
    \begin{align}
    \Bdisp=  & \argmin_{\B^T \M \B = \I}  \mathrm{tr}(\B^T \H \B). \label{eq:gevp-displacement-modes-derivation}
\end{align}
The optimal $\Bdisp$ may be found relatively efficiently with a generalized eigenvalue solver that supports large sparse matrices.
The columns of $\Bdisp$ can be directly interpreted as minimal-energy \emph{displacement} eigenmodes.

Our \emph{skinning} eigenmodes follows a similar derivation but we replace the optimization over $\B$ with an optimization over the weights $\W$. While \refeq{linear-blend-skinning-matmul} may appear to define $\Blbs$ as a complicated function of $\W$, it is \emph{linear} in and separable over the columns of $\W$. Thus, we may rewrite it as
 \begin{align}
     \Blbs &= \begin{bmatrix} \A_{i, j} & \dots & \A_{d, (d+1)} \end{bmatrix} (\I_{d(d+1)} \otimes \W),
     \label{eq:weight-space-skinning-jacobian}
\end{align}
where we introduce $\A_{i, j} \in \mathbb{R}^{n (d) \times n}$, our weight-space skinning Jacobians. These map contributions of each weight for all $d(d+1)$ affine parameters to the final skinning Jacobian. 

We derive $\A_{i, j}$ for the $d=2$ and $d=3$
in Appendix \ref{appendix-sec:weight-space-skinning-jacobians}, but for clarity show the result for $d=3$ here:
\begin{align}
\begin{matrix}
\A_{1, 1} = \P_x \bar{\X}  &\A_{1, 2} = \P_x \bar{\Y} &\A_{1, 3} = \P_x \bar{\Z} &
\A_{1, 4} = \P_x \\
\A_{2, 1} = \P_y \bar{\X} &\A_{2, 2} = \P_y  \bar{\Y} &\A_{2, 3} = \P_y \bar{\Z} &
\A_{2, 4} = \P_y \\
\A_{3, 1} = \P_z \bar{\X}  &\A_{3, 2} = \P_z \bar{\Y} &\A_{3, 3} = \P_z \bar{\Z} &
\A_{3, 4} = \P_z
\end{matrix}
\label{eq:weight-space-skinning-jacobian}
\end{align}
where the $\P_* \in \mathbb{R}^{3n \times n}$ selection matrices concatenate to form the identity matrix $\I_{3n} = [\P_x\, \P_y\, \P_z]$ and 
$\bar{\X}, \bar{\Y}, \bar{\Z} \in \mathbb{R}^{n\times n}$ are diagonal matrices containing the the $x$, $y$ and $z$ rest position values.




Following a similar procedure as before, we add a weight space orthogonality constraint $\W^T \M_{w} \W = \I$ and assume a generic distribution $\mathcal{D}$ on our sampling of $\z \sim \mathcal{D}$ to obtain

\edit{
 \begin{align}
     \W = &\argmin_{\W^T \M_{w} \W = \I} \mathrm{tr}(  \B_{\mathrm{lbs}}^T \H \B_{\mathrm{lbs}}\mathbb{E}_{\z \sim \mathcal{D}} [\z\z^T]) . 
 \end{align}}
 We now need to make assumptions on the distribution of $\z$, as these now correspond to flattened affine matrix parameters and so have some structure to their distribution.
 Specifically we assume that parameters belonging to different affine matrices are i.i.d. with respect to each other, but generally allow for intradependence between parameters belonging to the same affine matrix, as measured by the covariance matrix  $\C \in \mathbb{R}^{d(d+1) \times d(d+1) }$.
 \begin{align}
    \W = &\argmin_{\W^T \M_{w} \W = \I} \mathrm{tr}\left( (\I_{d(d+1)} \otimes \W)^T \A^T \H \A (\I_{d(d+1)} \otimes \W) (\I_m \otimes \C)\right).
    \nonumber
\end{align}
Expanding out all the Kronecker products and leveraging that the trace is just a sum of diagonal entries :
 \begin{align}
     \W = &\argmin_{\W^T \M_{w} \W = \I} \mathrm{tr}\left( \W^T \left(\sum_i^{d(d+1)} \sum_j^{d(d+1)} (\A^T_i \H \A_j)  c_{ij} \right) \W  \right).
     \end{align}
 Leading to the weight-space optimization:
 \begin{align}
    \W = &\argmin_{\W^T \M_{w} \W = \I} \mathrm{tr}\left( \W^T \H_{w} \W \right)
     \label{eq:gevp-skinning-modes-derivation}
 \end{align}
where $\M_{w} \in \mathbb{R}^{n \times n}$ is the weight-space mass matrix (we use the diagonal lumped mass matrix) and 
where we call $\H_{w} \in \mathbb{R}^{n \times n}$ the \emph{weight-space} elastic energy Hessian.

\begin{figure*}
\includegraphics[width=\textwidth, keepaspectratio]{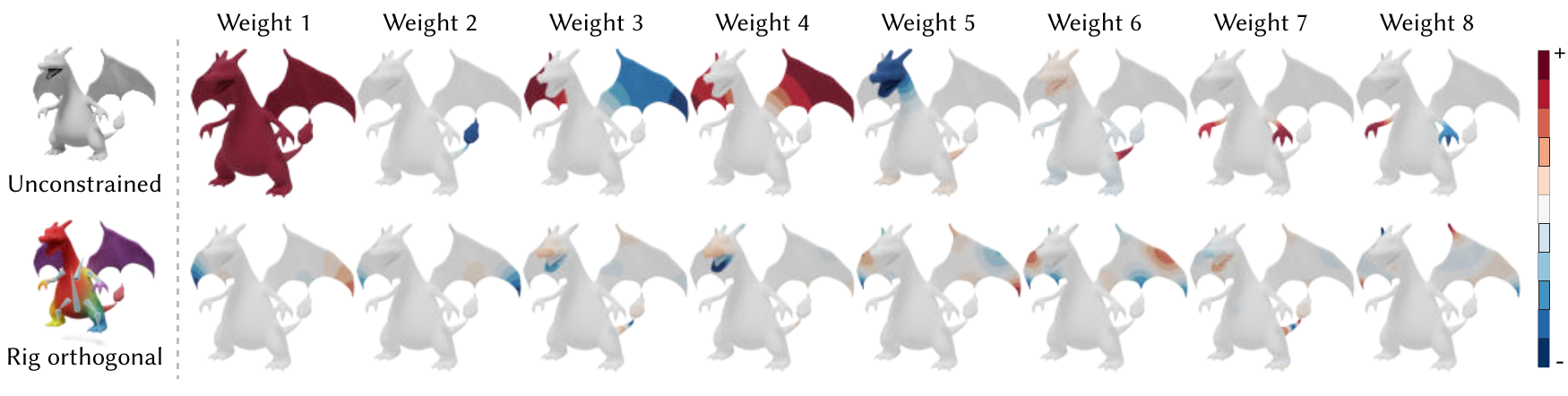}\timestamp[-0.25cm]{\tsWeightVis}
\caption{
We generate a linear blend skinning subspace for secondary motion parameterized by a set of skinning weights. Each weight $i$ shown is independently normalized to lie between $[-1, 1]\text{abs}(\boldsymbol{W}_i)$ and centered around 0. (Top) Weights generated by solving the unconstrained generalized eigenvalue problem on a weight-space elasticity Hessian. (Bottom) Secondary skinning weights that satisfy the weight-space complementarity constraint and are orthogonal to our rig space. These are naturally rig-aware, leading to higher frequency motion. \label{fig:skinning-weights-for-secondary-motion}}
\end{figure*}

\begin{figure}
\includegraphics[width=\linewidth,keepaspectratio]{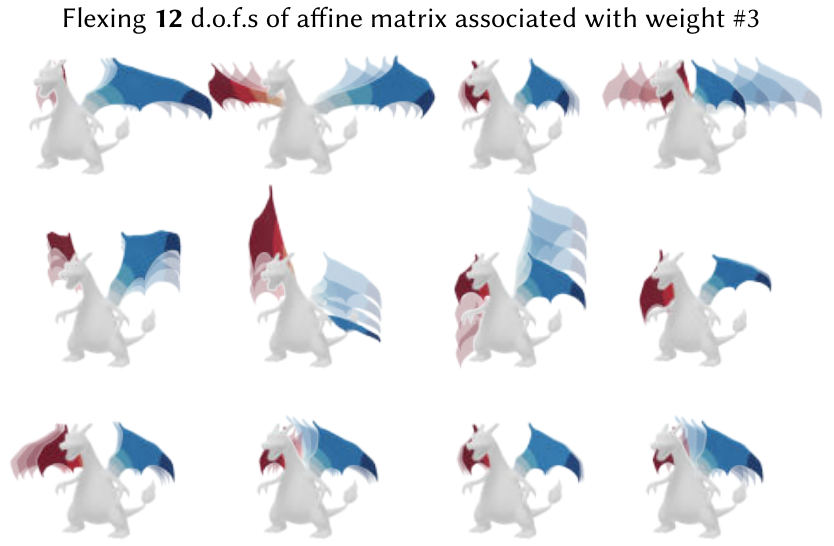}\timestamp{\tsAffineMotions}
\caption{
One secondary linear blend skinning weight could produce 12 different motions, corresponding to 12 d.o.f.s of an affine matrix. We showcase this by flexing those associated with weight \#3. \label{fig:motions-producible-by-skinning-modes}}
\end{figure}

\begin{figure}
\includegraphics[width=\linewidth,keepaspectratio]{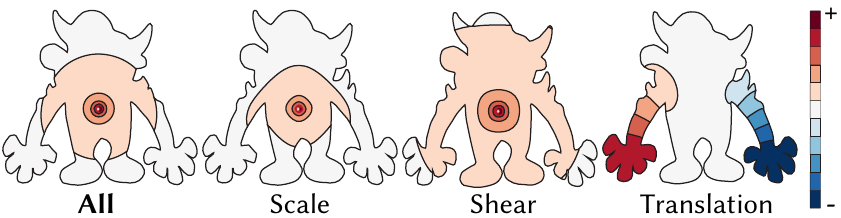}
\caption{ Prioritizing scaling and shearing (middle left and middle right) provides weights that are unnaturally centered around the origin. For this reason, we prioritize translations (right). \label{fig:prioritizing-affine-parameters-as-subspace-for-def}}
\end{figure}

\begin{figure}
\includegraphics[width=\linewidth,keepaspectratio]{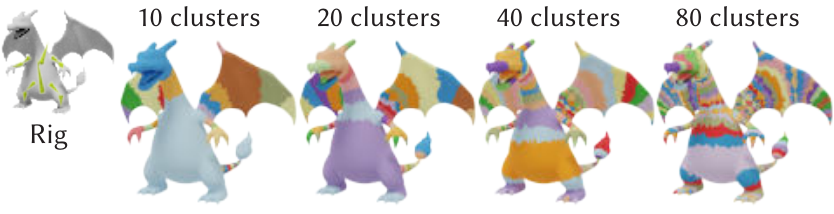}
\caption{
We generate clusters to accelerate the computation of per-tet energetic non-linearities. Our clusters inherit the rig-sensitivity of our skinnning weights. \label{fig:cluster-vis}}
\end{figure}

It is important to note this is overly determined for $\W$; The same set of weights are used to specify $d(d+1)$ different types of affine motions: scales, shears and translations. As a result, the set of weights that leads to optimal translations may not be the same set of weights that lead to optimal scales or shears. We can change which of these parameters we prioritize by modifying our affine parameter covariance matrix $\C$.

We choose to prioritize translations, neglecting shears and scales entirely, which are poorly suited for deformation subspaces.
The logic is that shears and scales are origin-dependent.
This leads the optimization in \refeq{gevp-skinning-modes-derivation} to see vertices far from the origin as \emph{stiffer} than vertices that are close to it, resulting in weights that are unnaturally concentrated around the origin, and decay far away from it as shown in \reffig{prioritizing-affine-parameters-as-subspace-for-def}. 

For $d=3$, taking i.i.d. samples from the standard normal distribution of each of the three translation parameters, while neglecting shears and scales leads to a covariance matrix of the form:
\begin{align}
    \C = \I_{3} \otimes 
    \begin{bmatrix} 
    1 & 0 & 0 & 0 \\
    0 & 0 & 0 & 0 \\
    0 & 0 & 0 & 0 \\
    0 & 0 & 0 & 0 \\
    \end{bmatrix}
\end{align}
which very conveniently leads to a simplified weight space Hessian:
\begin{align}
    \H_w = \P_x^T \H \P_x +  \P_y^T \H \P_y + \P_z^T \H \P_z.
    \label{eq:weight-space-hessian}
\end{align}

\begin{wrapfigure}{r}{5.0cm}
\includegraphics[width=\linewidth,keepaspectratio]{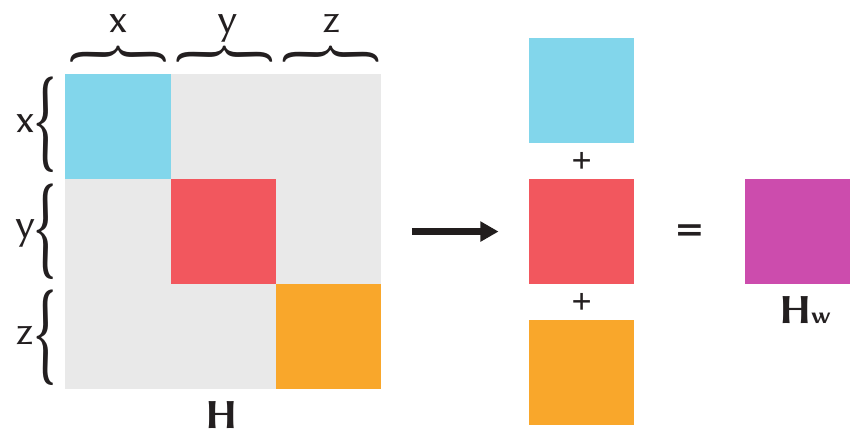}
\end{wrapfigure}
The inset, unburdened by notation, more clearly shows the simplicity of deriving this final weight-space Hessian; just take the diagonal blocks for each dimension of the Hessian and sum them up.
For co-rotational elasticity with homogeneous materials, $\H_w$ is proportional to the mesh's cotangent Laplacian matrix.
Whereas heterogeneous materials distributions affect $\H_w$ non-trivially and thus also the our optimal weights $\W$.

With these matrices defined, our optimal skinning eigenmodes are solutions to
\refeq{gevp-skinning-modes-derivation}, found efficiently via a genearlized eigenvalue solver.
%
%
Each individual skinning eigenmode --- as a linear blend skinning weight --- corresponds to $d(d+1)$ degrees of freedom and may be used to generate  $d(d + 1)$ different motions, as shown in \reffig{motions-producible-by-skinning-modes} for $d=3$.
%

\subsubsection{Weight Space Complementarity Constraint}
At run-time, our secondary-effect displacements should satisfy $\J^T \u^c = \boldsymbol{0}$, where recall $\J \in \mathbb{R}^{3n \times \dimp}$ is the current rig Jacobian.
In our subspace, this becomes $\J^T \Blbs \z = \boldsymbol{0}$.
Without knowledge of $\J$ \emph{a priori}, our optimized skinning eigenmodes will, in general, not admit non-trivial solutions. Even if they did, enforcing this constraint at run-time leads to a more difficult constrained optimization problem.
Fortunately, our formulation above as a generalized eigenvalue problem allows us to 
easily add constraints to our skinning weights, thus ensuring that our modes admit non-trivial solutions but also implicitly satisfy the constraint allowing us to remove it entirely at run-time.

\citet{Zhang:CompDynamics:2020} define $f_\text{rig}(\p)$ generically. For real-time  applications, we will assume that $f_\text{rig}$ is linear (single affine handle, linear blend skinning, blendshapes, etc.) and thus has a constant rig Jacobian $\J$.
Given $\J$, the constraint we need to add is
\begin{align}
    \J^T \Blbs = \boldsymbol{0} 
\end{align}

To express this in terms of $\W$, we can again make use of our weight-space skinning Jacobian matrices from \refeq{weight-space-skinning-jacobian} (not to be confused with $\J$) and expand the constraint to act on each weight.
This leads to a series of constraints that our weights need to satisfy: 
\begin{align}
\begin{matrix}
\J^T \A_{i, j} \W = \boldsymbol{0}
\end{matrix}  \in \mathbb{R}^{\dimp \times \nummodes}
\quad \forall i \in \{1, ...,d\}, \, j  \in \{1, ..., d+1\}
\end{align}
We can stack all our constraint matrices $\J^T \A_{i, j}$:
\begin{align}
\begin{bmatrix}
\J^T\A_{1, 1} \\
\vdots \\
\J^T \A_{d, d+1}\\
\end{bmatrix} 
\W = \J_w \W = \boldsymbol{0} \in \mathbb{R}^{p(d)(d+1) \times m}
\label{eq:weight-space-complementary-constraint}
\end{align}
where we call $\J_w \in \mathbb{R}^{ \dimp (d)(d+1) \times n}$ our weight-space complementarity constraint matrix.

We can incorporate this constraint in a standard generalized eigenvalue problem by solving instead
\begin{align}
    \begin{bmatrix}
    \H_{w}  & \J_{w}^T \\
    \J_{w} & \boldsymbol{0}
    \end{bmatrix}
    \begin{bmatrix}
    \W \\
    \boldsymbol{\mu}
    \end{bmatrix}
    =
     \begin{bmatrix}
    \M_{w}  & \boldsymbol{0} \\
    \boldsymbol{0} & \boldsymbol{0}
    \end{bmatrix} 
        \begin{bmatrix}
    \W \\
    \boldsymbol{\mu}
    \end{bmatrix}
    \boldsymbol{\Lambda}.  
    \label{eq:gevp-skinning-modes-constrained}
\end{align}

\reffig{skinning-weights-for-secondary-motion} shows how our derived skinning weights change to accommodate the rig-complementarity constraint.



\begin{figure*}
\includegraphics[width=\linewidth,keepaspectratio]{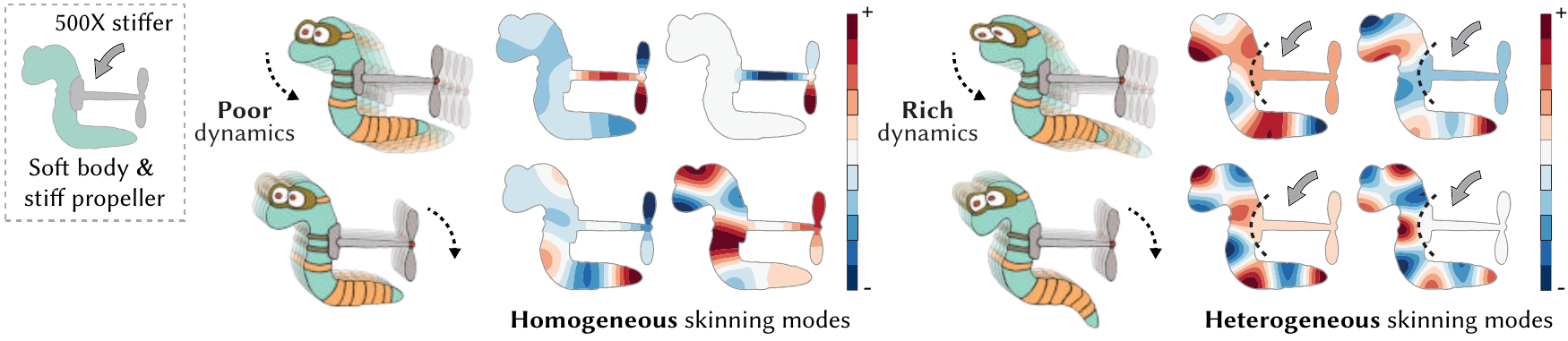} \timestamp[-0.25cm]{\tsHeterogeneousMaterialExperiment}
\caption{A material-aware subspace more efficiently captures the space of motions available to our simulation. This directly leads to richer dynamics.\label{fig:heterogeneous-skinning-modes}}
\end{figure*}

\section{But What About Rotations?}
\label{sec:rotations}
Interesting elastodynamic effects exhibit rotations: both global, where the entire shape rotates in space, and local, where part or parts of the shape rotate relative to the rest/each other.
Rotations are notoriously difficult for previous linear subspaces.
For example, it is well known that displacement modes ($\Bdisp$ defined in \refeq{gevp-displacement-modes-derivation}) 
struggle to represent local rotational motion (see \reffig{local-rotation-experiment} and \reffig{rotation-fitting} (Left)) \cite{BarbicJames:RealTimeSTVK,Barbic:2011:RealTimeLargeDefoSubstructuring, RScoords, ModalWarping}). 

Our use case reveals that issues with rotations go beyond this and can be more insidious.
In the following discussion, we assume that the elastic potential $E$ is rotation invariant. That is, $E(\u + \x_0) = E(\repR (\u + \x_0))$, where multiplying by $\repR \in \mathbb{R}^{n(d) \times n(d)}$ applies the same rotation $\R \in SO(d)$ to all vertices.

\subsection{Rotation Spanning vs. Closure Under Rotations}
Rotational problems may be categorized into two separate issues.

First, does a given subspace span rotations?
By global rotation spanning, we mean there always exists some subspace parameters to reproduce any rotational displacement. If $\x_0$ are the rest positions then 
\begin{align}
\exists \ \z  \text{ such that } \repR \x_0 - \x_0 = \B \z \ \forall \R \in SO(d).
\end{align}
For free-flight objects, failing to span global rotations means the subspace will unnaturally deform in an attempt to minimize $E$ rather than rotate.
%
%
%
Unfortunately, this problem also occurs locally, too. For example, if the arms of a character bend in opposite directions, failure to span these local rotations will disturb (by introducing local shears and stretches to attempt to minimize $E$) or prevent (by the minimization of $E$ detesting such scales and shears) the desired deformation.
For example, \citet{BarbicJames:RealTimeSTVK} emphasize how displacement modes $\Bdisp$ induce scaling and shearing artifacts when approximating rotations and bending deformations.

Second, does a given subspace induce a rotationally equivariant simulation?
Treat the simulation as a map from problem specification parameters (e.g., forces, rig displacements, rest positions) to optimal (full-space) displacements.
Rotation equivariance means that any rotated version of the problem results in a correspondingly rotated solution:
\begin{align}
\forall \boldsymbol{R} \in \mathcal{SO}(d)\, ,  & \quad 
\forall \boldsymbol{x} \in \mathbb{R}^{n(d)}\, , \nonumber\\
     \boldsymbol{B} \argmin_{\boldsymbol{z} }{\boldsymbol{E}(\boldsymbol{B}\boldsymbol{z} + \repR \boldsymbol{x}}) &= 
    \repR \boldsymbol{B} \argmin_{\boldsymbol{z} }{\boldsymbol{E}(\boldsymbol{B}\boldsymbol{z} + \boldsymbol{x}}) \ ,
    \label{eq:rotation-equivariance-of-sim}
\end{align}
where --- without loss of generality --- we lump problem specification parameters into the vector $\x \in \mathbb{R}^{n(d)}$.

A subspace simulation lacking rotation equivariance may experience unpredictably different deformations under rotations.
This is especially problematic in a complementary dynamics setting where the entire object or large subpart may rotate due to the user rig. Users will expect analagously rotated secondary effects and be surprised by behavior that depends on global or local rotations coming from the rig.

 \begin{wrapfigure}[13]{r}{3.5cm}
\includegraphics[width=3.8cm,keepaspectratio]{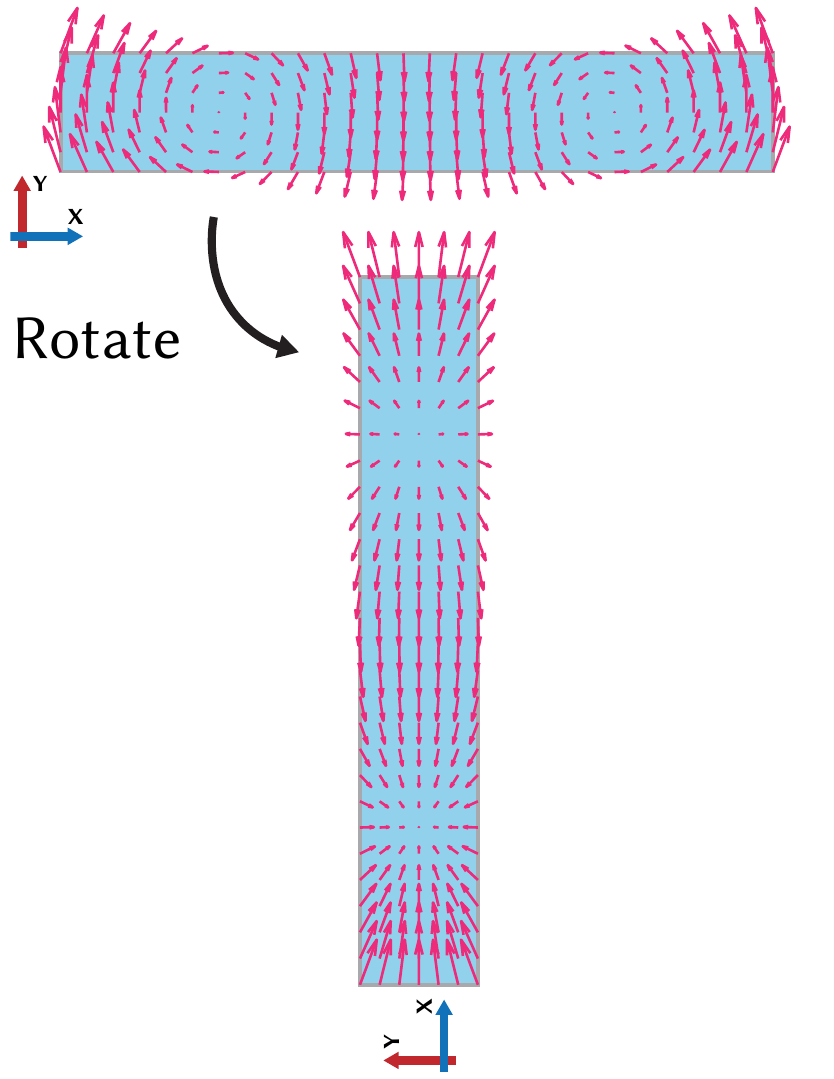}\timestamp{\tsBarFieldVis}
\end{wrapfigure}
\reffig{subspace-comparisons} shows how this shortcoming expresses itself as  overly energetic deformation, while \reffig{random-init-linear-subspace} showcases some of the kinky local minima that can easily arise under simple rotations. 
The root of this problem is shown didactically in the inset: a single displacement mode describes a completely different type of motion if its underlying shape rotates.

Building on this intuition, we prove (see App.~\ref{sec:appendix-proof-sim-rotation-equivariance}) that a linear subspace simulation is rotation equivariant if and only if the subspace basis is \emph{closed under rotations}:
\begin{align}
    \forall \, \boldsymbol{R}\in \mathcal{SO}(d) \text{ and } \boldsymbol{z}\in\mathbb{R}^m \ \exists \, \boldsymbol{w}\in\mathbb{R}^m, \text{ such that }   \repR \boldsymbol{Bz} =  \boldsymbol{B} \boldsymbol{w}.
    \label{eq:rotation-equivariance-requirement} 
\end{align}

%


\subsection{Displacement Modes Simulations Are Fragile Under Rotations}
Displacement modes ($\Bdisp$ defined in \refeq{gevp-displacement-modes-derivation}) \cite{PentlandWilliams1989} and many of their 
improvements (e.g., \cite{BarbicJames:RealTimeSTVK}) are neither rotation spanning nor closed under rotations.
Rotations are a full-spectrum displacement, so any (reasonable) truncated elastic eigenspace will fail to span arbitrary global rotations (see \reffig{rotation-fitting} (Left)).
While --- as discussed above --- global rotation spanning has an easy fix, much effort has been made to improve local rotation spanning such as data-driven PCA bases \cite{EigenSkinKry2022}, modal derivatives
\cite{BarbicJames:RealTimeSTVK}, sub-structuring \cite{Barbic:2011:RealTimeLargeDefoSubstructuring}, or splitting the simulation into rigid and deformable components \cite{Terzopoulos1988}.

\begin{figure}
\includegraphics[width=\linewidth,keepaspectratio]{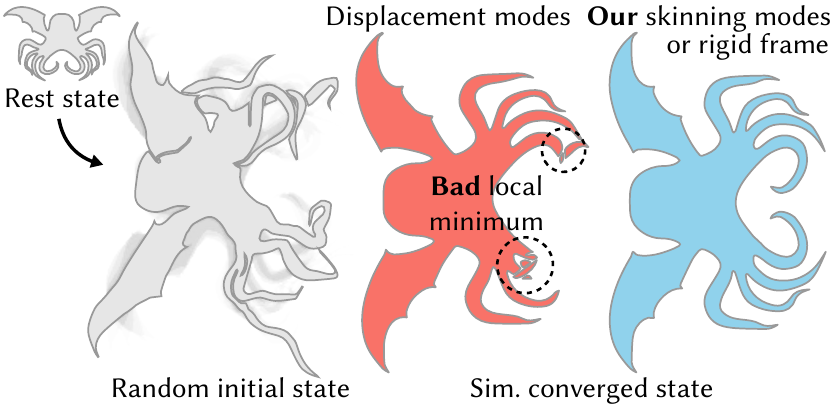}\timestamp{\tsCthuluLocalMinimum}
\caption{
We compute the subspace at rest (top-left). A user rotates the mesh and perturbs the system with a random initial deformation.  Using displacement modes creates jarring local minimum artefacts in a rotated frame. Our skinning modes find the global minimum effortlessly, \edit{obtaining the same rest state than if we had embedded the simulation in a rigid frame. \label{fig:random-init-linear-subspace}}}
\end{figure}

%
Nevertheless, large local rotations may still be problematic (see \reffig{local-rotation-experiment}).
Displacement modes --- except if truncated to just null modes or completely non-truncated --- are not closed under rotations (see counterexamples in \reffig{rotation-fitting} (Right) and \reffig{random-init-linear-subspace}).
%


\subsection{Skinning Eigenmodes Are Robust to Rotations}
\label{sec:skinning-modes}
\begin{figure}
\includegraphics[width=\linewidth,keepaspectratio]{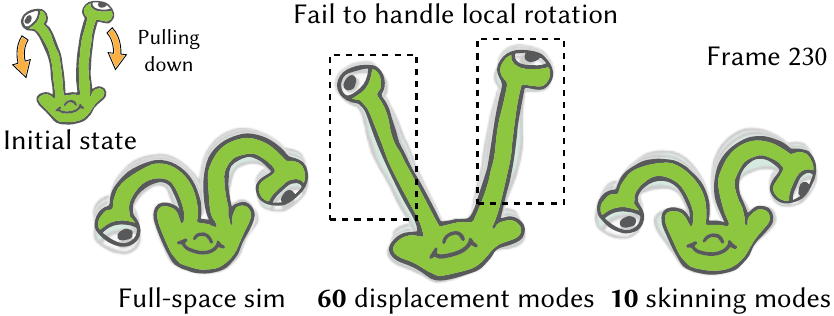}\timestamp[-0.125cm]{\tsBingbyLocalRotation}
\caption{No matter how hard a user tries, the eyes of this reduced elastodynamic alien will never bend when using a small displacement mode subspace (middle). Our skinning subspace (right) enables the rotational motion. Both results use 60 degrees of freedom. 
\label{fig:local-rotation-experiment}} 
\end{figure}

\begin{figure*}
\includegraphics[width=\linewidth,keepaspectratio]{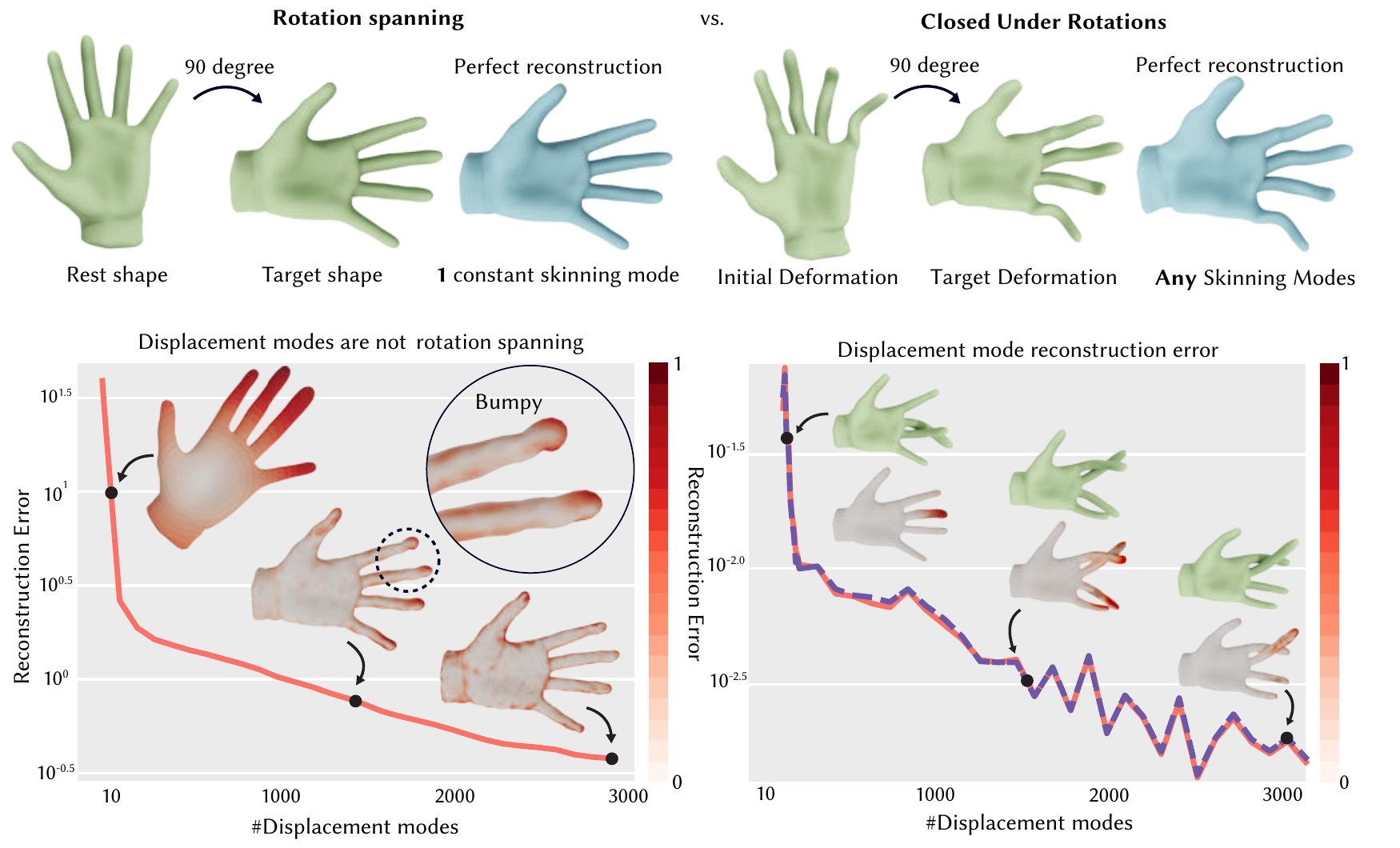}
\caption{ \textbf{Rotation Spanning vs. Closure Under Rotations.} Given some initial shape, we optimize for optimal displacements that minimize the squared distance between each vertex position and its rotated target. (Left) Our skinning modes are \textbf{rotation spanning}, as modulated by the skinning weights. With a single constant skinning weight, we can perfectly reconstruct (by least squares projection) any rotation of the rest shape. Displacement modes do not span rotational motion, even with excessively abundant modes. (Right-red) Our skinning modes are \textbf{closed under rotations}, so any deformation in the span of those can be reconstructed (by least squares projection) under the same set of modes even if the mesh is arbitrarily rotated by a user. \edit{The same cannot be said for even an impractically large number of displacement modes even if augmented with affine degrees of freedom  (right-purple).}
 \label{fig:rotation-fitting}}
\end{figure*}

In contrast, skinning eigenmodes are both rotation spanning and closed under rotations (see \reffig{rotation-fitting}).
When the complementarity constraint is absent, the first skinning eigenmode will be a constant function thus spanning all affine motions including rotations.
When used for fast complementary dynamics, the rig typically contains global rotations so we explicitly (and purposefully) avoid global rotation spanning.
We do still want and indeed observe local rotation spanning (see \reffig{local-rotation-experiment}).
%
%
%
%
We can easily show that the linear blend skinning --- and thus also skinning subspaces --- are closed under rotations. 
Given some rotation $\boldsymbol{R} \in \mathcal{SO}(d)$, rotating linear blend skinning's output is equivalent to rotating all of the input transformations:
\begin{equation}
\R \sum_{b=1}^m w_{ib} \T_b \X_i = 
\sum_{b=1}^m w_{ib} \R \T_b \X_i.
    \label{eq:linear-blend-skinning}
\end{equation}
Any rotation of its output is producible by its input, as required for a rotation equivariant subspace simulation.
This fact was similarly utilized in previous works albeit in different settings \cite{Wang:2015:LinearSubspaceDesign,JacobsonBPS11,LangerS08}.
%
\edit{
There are many prior  subspaces \cite{Faure2011, 1Gilles2011, Wang:2015:LinearSubspaceDesign, Tycowicz2013}  that do not explicitly mention rotation equivariance as a criterion for the subspace simulation. In hindsight, leveraging the machinery of Appendix \ref{appenix-eq:closed-under-rotations}, we can see that since these prior subspaces are closed under rotations, those methods also maintain a rotation equivariant simulation.
}

\section{Clustering}
\label{sec:clustering}
For many rich rotational dynamic properties, an elastic energy usually requires the use of per-tetrahedron non-linear operations \cite{KimDynamicDeformables}. Despite our use of a subspace for our displacement degrees of freedom, these energies \emph{still} demand per-tet computation. Our simulations are not yet truly decoupled from the resolution of the mesh.

Many methods in the past sidestep this issue by making use of \emph{cubature} \cite{BarbicJames:RealTimeSTVK,TengArticulatedBodyContact,OptimizingCubature}, where the deformation quantities at each tetrahedra are estimated as a sum of weighed contributions from a sparse set of pre-determined sample tetrahedra. These samples and their interpolation weight are usually obtained through a data-fitting phase. 
We instead opt for a clustering \cite{Jacobson-12-FAST} scheme, which estimates these non-linearities through $r$ clusters of tetrahedra. Instead of requiring training data, we compute our clusters via a deformation prior, obtaining them via a $k$-means clustering \cite{kmeanspp} of our subspace secondary skinning weights $\boldsymbol{W}$.
For a mesh with $\numtets$ tetrahedra, we use $\numclusters$ clusters
  to build the grouping matrix $\boldsymbol{G} \in \mathbb{R}^{\numclusters  \times  \numtets}$:
\begin{align}
\boldsymbol{G}_{ij}  =
 \begin{cases}
  \frac{\boldsymbol{V_j}}{\sum_q^{\mathcal{C}_i}{\boldsymbol{V_q}}}  \quad & \text{if} j \in \mathcal{C}_i \\
  0  \quad & \text{otherwise}
 \end{cases}
\end{align}
where $\boldsymbol{V}_j$ is the volume of the $j$-th tetrahedron, while $\mathcal{C}_i$ contains the indices of the tetrahedra belonging to cluster $i$.
This grouping matrix computes cluster quantities via a mass weighed averaging of its member tetrahedra.  We use these grouping matrices whenever we require per-tet evaluations in our solver as outlined in \refsec{local-global-solver}.

We build our clusters via a $k$-means++ \cite{kmeanspp}, using features obtained from our secondary weights. Specifically, we average our vertex weights onto each tetrahedra, then we scale each of our weights $\boldsymbol{w}_i$ by the inverse-squared of its associated eigenvalue $\lambda_i$, obtained in \refeq{gevp-skinning-modes-constrained}. 
\reffig{cluster-vis} shows clustering results for a sample mesh with different rigs. Because our skinning subspace is the only feature we use for building our clusters, our clusters inherit many of the properties of the skinning modes, including being material-sensitive, rig-sensitive. This allows our clusters to provide a higher resolution in parts of the mesh more likely to exhibit secondary motion.
Because our skinning modes can be global, this results in clusters that could also be global. We detect these in a post processing step and \emph{split} clusters into their individual separate components. 

\section{Local-Global Solver}
\label{sec:local-global-solver}
We make use of a simple local-global solver for the minimization of our elastodynamic energy. This solver leverages our reduced space degrees of freedom and our clusters to \emph{never} make use of any full-space operations. This solver allows us to avoid recomputing expensive quadratic energy Hessians each timestep while still accommodating frequently desired elastic non-linearities, such as rotations. 

\subsection{Full-Space Local-Global Solver}
We assume energies that can be decomposed as
\begin{equation}
    E(\boldsymbol{u}) = \Psi(\boldsymbol{u}) + \Phi(\boldsymbol{u}) \, , \nonumber
\end{equation}
where $\Psi(\boldsymbol{u})$ is quadratic in $\boldsymbol{u}$   and $\Phi$ is a non-linear in $\boldsymbol{u}$ and piecewize constant over tets. We then need to identify the source of the non-linearities in $\Phi$, $\boldsymbol{R}$, as auxiliary degrees of freedom
\begin{equation}
  \boldsymbol{u}, \boldsymbol{R} =  \argmin_{\boldsymbol{u}, \boldsymbol{R}}  \Psi(\boldsymbol{u}) +  \Phi(\boldsymbol{u}, \boldsymbol{R}) \, . \nonumber
\end{equation}


We split elastodynamic optimization into two steps  \cite{ARAPolgaSorkine}. The 
local step optimizes for the non-linear sources $\boldsymbol{R}$ in our energy: 
holding the primary degrees of freedom fixed, \begin{align}
  \boldsymbol{R}_{i} &=  \argmin_{\boldsymbol{R}}  \Phi(\boldsymbol{u}_{i-1}, \boldsymbol{R}) \, .\label{eq:full-space-local-step}
 \end{align}
The global step optimizes for the primary degrees of freedom, holding the auxillary ones fixed:
\begin{align}
  \boldsymbol{u}_{i} &=  \argmin_{\boldsymbol{u}}  \Psi(\boldsymbol{u}) +  \Phi(\boldsymbol{u}, \boldsymbol{R}_{i})\, .
  \label{eq:full-space-global-step}
\end{align}
These two steps are repeated iteratively until convergence..

\begin{figure}
\includegraphics[width=\linewidth,keepaspectratio]{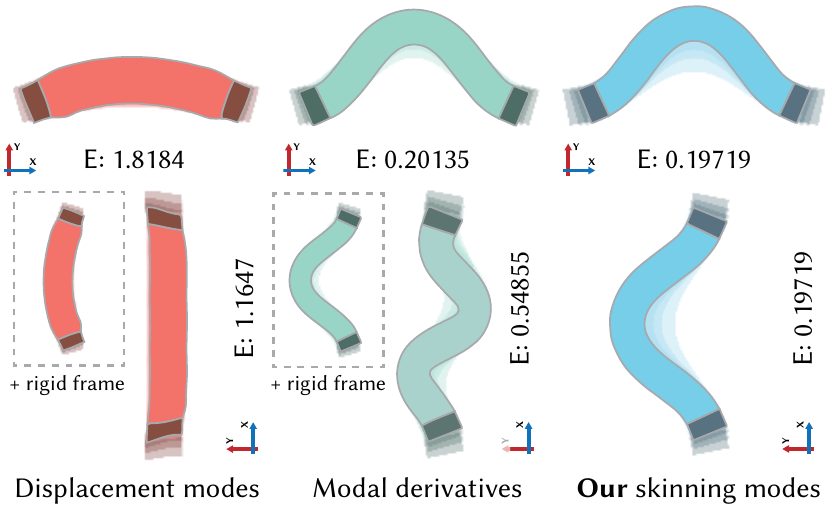}\timestamp[-0.125cm]{\tsSubspaceComparison}
\caption{
Simulations making use of displacement modes (left) are not rotation equivariant. Augmenting displacement modes with modal derivatives (middle), helps in representing higher quality deformations, but still create different high energy deformations under rotations. Our skinning modes (right)  allow us to capture both local rotational motion and maintain rotation equivariance \edit{and lead to the lowest energy even without explicitly tracking a rotating frame  \cite{Terzopoulos1988} each timestep (bottom)}. } \label{fig:subspace-comparisons}
\end{figure}

\subsection{Hyper-Reduced Local-Global Solver}
To ensure our solver never performs any full space operations, we make use of our subspace $\boldsymbol{u} \approx \boldsymbol{B} \boldsymbol{z}$ and our clusters  $\Phi \approx \tilde{\Phi}$, where the tilde denotes clustered quantities. While before $\boldsymbol{\Phi}$ was piecewise constant over \emph{tets}, it is now piecewise constant over \emph{clusters}, where clustered physical quantities, (like mass, Lamé parameters, deformation gradients, etc.) have been averaged over all the tets in each cluster. 
Our hyper-reduced energy becomes:
\begin{equation}
    E(\boldsymbol{u}) = \Psi(\boldsymbol{z}) + \tilde{\Phi}(\boldsymbol{z}) \, .\nonumber
\end{equation}
Exposing the per-cluster source of non-linearities as their own auxillary degrees of freedom $\boldsymbol{\tilde{R}}$:
\begin{equation}
    E(\boldsymbol{z}) = \Psi(\boldsymbol{z}) + \tilde{\Phi}(\boldsymbol{z}, \boldsymbol{\tilde{R}}) \nonumber
\end{equation}
Our hyper-reduced local step becomes
\begin{align}
  \boldsymbol{\tilde{R}}_{i} &=  \argmin_{\boldsymbol{\tilde{R}}}  \tilde{\Phi}(\boldsymbol{z}_{i-1}, \boldsymbol{\tilde{R}}) \, ,\label{eq:hyper-reduced-local-step}
\end{align}
\edit{Because the non-linearity is piece-wise constant over clusters, this per-cluster optimization can additionally be solved in parallel over all clusters, providing further acceleration.} 
while the hyper-reduced global step becomes
\begin{align}
  \boldsymbol{z}_{i} &=  \argmin_{\boldsymbol{z}}  \Psi(\boldsymbol{z}) +  \tilde{\Phi}(\boldsymbol{z}, \boldsymbol{\tilde{R}}_{i})\, .
  \label{eq:hyper-reduced-global-step} 
\end{align}

Depending on the specific energy, different optimization schemes need to be adopted for both the local step and the global step. The only criteria we need to maintain on any of these solvers is that they \emph{never} perform any full-space operations. Appendix \refsec{appendix-hyper-reduced-arap-elasticity} and \refsec{appendix-hyper-reduced-corotational-elasticity} show how to implement our reduced local-global solver for elastic energies such as ARAP \cite{ChaoGomElastic} and Linear Co-Rotational Elasticity \cite{McAdams2011}.

\section{Implementation}
Most of our implementation is in C++ with some calls to Matlab routines for building the subspace.  All timings reported were performed on a Dell XPS i9-12900HK with a 2.90 GHz processor and 64GB of RAM, equipped with an NVIDIA GeForce 3050Ti graphics card.

\begin{algorithm}[]
   \caption{Subspace Construction : given rest positions $\boldsymbol{X}$, tet indeces $ \boldsymbol{T}$, an elastic energy hessian $\boldsymbol{H}$ a complementarity constraint matrix $\mathcal{J}$, compute a subspace composed of $\mathrm{m}$ skinning weights $\boldsymbol{W}$ and  $\mathrm{r}$ cluster labels $\boldsymbol{l}$ for our tets. \label{alg:skinningSubspace}}

 \SetKwFunction{FcomputeSubspace}{computeSubspace}
  \SetKwProg{Fn}{Function}{:}{}
\Fn{\FcomputeSubspace{$\boldsymbol{X}, \boldsymbol{T}, \boldsymbol{H}, \mathcal{J}, \mathrm{m}, \mathrm{r}$}}{
$\boldsymbol{H}_w \gets \mathrm{weightSpaceHessian}(\boldsymbol{H})$
\\
$\boldsymbol{M}_w \gets  \mathrm{massmatrix(\boldsymbol{X}, \boldsymbol{T})}$
\\
$\boldsymbol{J}_w \gets  \mathrm{weightSpaceConstraint}(\mathcal{J}, \boldsymbol{X})$
\\
$
\mathcal{H} \gets
\begin{bmatrix} 
\boldsymbol{H}_w & \boldsymbol{J}^T_w \\
 \boldsymbol{J}_w & \boldsymbol{0} 
\end{bmatrix}
\; \mathcal{M} \gets 
\begin{bmatrix} 
\boldsymbol{M}_w & \boldsymbol{0} \\
\boldsymbol{0} & \boldsymbol{0}
\end{bmatrix}
$
\\
$\boldsymbol{W}, \boldsymbol{V} \gets \mathrm{eigs}(\mathcal{H}, \mathcal{M}, \mathrm{m})$
\\
$\boldsymbol{W_t}  \gets \mathrm{averageOntoTets}(\boldsymbol{W} \boldsymbol{V}^{-2}, \boldsymbol{T})$
\\
$\boldsymbol{l}  \gets \mathrm{kmeans}(\boldsymbol{W}_t, \mathrm{r})$
\\
$\boldsymbol{l}  \gets \mathrm{splitComponents}(\boldsymbol{T}, \mathrm{l})$
\\
\KwRet $\boldsymbol{W}, \boldsymbol{l}$
}

 \SetKwFunction{FConstraint}{weightSpaceConstraint}
  \SetKwProg{Fn}{Function}{:}{}
  \Fn{\FConstraint{$\mathcal{J}, \boldsymbol{X}$}}{
  $\boldsymbol{A} \gets 
[]$
\\
\For{$i\gets0$ \KwTo $\mathrm{dim}$}{
\For{$j \gets0$ \KwTo $\mathrm{dim}+1$}{
$\boldsymbol{A}_{i, j} \gets \mathrm{weightSkinningJacobian}(\boldsymbol{X}, i, j) \quad  \refeq{weight-space-complementary-constraint} $
\\
$\boldsymbol{A} \gets \begin{bmatrix}{} 	\boldsymbol{A}\\ 
	\mathcal{J}^T \boldsymbol{A}_{i, j} 
\end{bmatrix} $
}
}
$\boldsymbol{J}_w \gets \mathrm{removeRedundantRows} (\boldsymbol{A})$  
\\
\KwRet $\boldsymbol{J}_w$
  }
\end{algorithm} 
\begin{algorithm}[]
   \caption{performs one simulation step of our reduced Complementary Dynamics  using a linear Co-Rotated elasticity model \label{alg:simulationStep}}
 \SetKwFunction{FsimulationStep}{simulationStep}
  \SetKwProg{Fn}{Function}{:}{}
\Fn{\FsimulationStep{$\boldsymbol{p},\boldsymbol{z}_{hist}, \boldsymbol{p}_{hist},   \boldsymbol{f}_{ext}$}}{
$\boldsymbol{z} \gets \boldsymbol{0}$
\\
\While{$\mathrm{not\, converged}$}
{ 
$ \frac{\partial \boldsymbol{\tilde{\phi}}}{\partial \boldsymbol{f}} \gets \mathrm{localStep}(\boldsymbol{z}, \boldsymbol{p})$
\\
$ \boldsymbol{z}_{next} \gets \mathrm{globalStep}(\boldsymbol{z}, \boldsymbol{p}, \boldsymbol{z}_{hist}, \boldsymbol{p}_{hist}, \boldsymbol{f}_{ext}, \frac{\partial \boldsymbol{\tilde{\phi}}}{\partial \boldsymbol{f}})$
\\
$ \boldsymbol{z} \gets  \boldsymbol{z}_{next} $
}
\KwRet $\boldsymbol{z}_{next}$
}
 \SetKwFunction{FlocalStep}{localStep}
  \SetKwProg{Fn}{Function}{:}{}
\Fn{\FlocalStep{$\boldsymbol{z}, \boldsymbol{p}$}}{
$\boldsymbol{\tilde{f}} \gets \staticp{\boldsymbol{G_9K B}} \boldsymbol{z} + \staticp{\boldsymbol{G_9KJ_0}}\dynamicp{\boldsymbol{p}}$
\\
$ \boldsymbol{\tilde{F}} \gets \mathrm{reshape}(\boldsymbol{\tilde{f}}, [\mathrm{r}, \mathrm{dim} , \mathrm{dim} ])$
\\
$\frac{\partial \boldsymbol{\tilde{\phi}}}{\partial \boldsymbol{\tilde{F}}} \gets \mathrm{zeros}(\mathrm r, \mathrm{dim}, \mathrm{dim})$
\\
\For{$i\gets0$ \KwTo $\mathrm{r}$}
{
$\boldsymbol{F}_i \gets \boldsymbol{F}[i, :, :]$
\\
$ \boldsymbol{R}_i \gets \mathrm{PolarSVD}(\boldsymbol{F}_i)$
\\
$ \frac{\partial \boldsymbol{\tilde{\phi}}}{\partial \boldsymbol{\tilde{F}}}[i] \gets -m_{i} \mu_{i} \boldsymbol{R}_{i} + m_{i} \frac{\lambda_i}{2} \boldsymbol{R}_i tr(\boldsymbol{R}_i^T \boldsymbol{F}_i - \boldsymbol{I})$
\\
}
$ \frac{\partial \boldsymbol{\tilde{\phi}}}{\partial \boldsymbol{f}} \gets \mathrm{vec}(\frac{\partial \boldsymbol{\tilde{\phi}}}{\partial \boldsymbol{\tilde{F}}})$
\\
\KwRet $ \frac{\partial \boldsymbol{\tilde{\phi}}}{\partial \boldsymbol{f}}$
}

 \SetKwFunction{FglobalStep}{globalStep}
  \SetKwProg{Fn}{Function}{:}{}
\Fn{\FglobalStep{$\boldsymbol{z},\boldsymbol{z}_{hist}, \boldsymbol{p}, \boldsymbol{p}_{hist},   \boldsymbol{f}_{ext}, \frac{\partial \boldsymbol{\tilde{\phi}}}{\partial \boldsymbol{f}})$}}{
$\boldsymbol{f}_{elastic} \gets \staticp{\boldsymbol{B}^T \boldsymbol{LJ}} \dynamicp{\boldsymbol{p}} + (\staticp{\boldsymbol{G_9KB}})^T \frac{\partial \boldsymbol{\tilde{\phi}}}{\partial \boldsymbol{f}}  $
\\
$\boldsymbol{f}_{inertia} \gets \frac{1}{h^2}( \staticp{\boldsymbol{B}^T \boldsymbol{MJ}}\dynamicp{(\boldsymbol{p} - \boldsymbol{p}_{hist})} - \staticp{\boldsymbol{B}^T \boldsymbol{MB}} \dynamicp{\boldsymbol{z}_{hist}}) $
\\
$\boldsymbol{f} \gets  \boldsymbol{f}_{elastic} +  \boldsymbol{f}_{inertia} + \boldsymbol{f}_{ext}$
\\
$\boldsymbol{A} \gets \staticp{\boldsymbol{B}^T\boldsymbol{L}\boldsymbol{B}} + \frac{1}{h^2} \staticp{\boldsymbol{B}^T\boldsymbol{M}\boldsymbol{B}}$
\\
$ d\boldsymbol{z} \gets \boldsymbol{A} d\boldsymbol{z} = - \boldsymbol{f} - \boldsymbol{A}\boldsymbol{z} $
\\
$\alpha \gets \mathrm{lineSearch}(d\boldsymbol{z})$ 
\\
$\boldsymbol{z}_{next} \gets \boldsymbol{z} + \alpha d\boldsymbol{z} $
\\
\KwRet $\boldsymbol{z}_{next}$
}
\end{algorithm} 

\subsection{Subspace Construction}
Algorithm \ref{alg:skinningSubspace} provides pseudocode for the construction of our subspace. We require as input the rest state, an elastic energy Hessian, and a user-defined homogeneous linear equality constraint matrix $\mathcal{J}$, in our case the complementarity constraint matrix.
So far, for conciseness, we have stated that our complementarity  constraint matrix was equal to the rig jacobian $\mathcal{J} = \boldsymbol{J}$. However, as described by \cite{Zhang:CompDynamics:2020}, to make our complementarity constraint sensitive to the mesh resolution, it is also weighed by a mass-matrix.
Additionally the constraint matrix is weighed by a momentum-leaking matrix $\boldsymbol{D}$. This diagonal matrix with entries ranging form $0$ to $1$ specifies where momentum can leak from the rig to the mesh. By default, we compute these diagonal entries from a fast diffusion of the surface to the interior, and renormalize the values to lie between $0$ and $1$, which easily gives us standard \emph{follow-through} and \textit{anticipation} effects \cite{Zhang:CompDynamics:2020}. For additional control, a user can also customize and easily provide their own scalar field designed by a method of their choice. The final complementarity constraint matrix is $\mathcal{J}= \boldsymbol{D} \boldsymbol{M} \boldsymbol{J}$.

For the GEVP solver, we use Matlab's $\mathrm{eigs}$. We also remove redundant rows from our weight space complementarity constraint quickly using Matlab's $\mathrm{rref}$.
\subsection{Simulation Step}
Algorithm \ref{alg:simulationStep} provides an overview for the simulation step of our proposed method.
We color in \staticp{\staticColorName} matrix products that can be computed \emph{once} at the start of the simulation, cached and then called at run-time. We color in \dynamicp{\dynamicColorName} vectors that do not change throughout the multiple iterations of a single local-global solve. Products with these red vectors can be computed once at the start of the timestep step and reused throughout the timestep.
To solve the linear system for the Quasi-Newton search direction, we make use of a precomputed Cholesky Factorization. In the case of vanishing Poisson ratio, where Co-Rotational Elasticity becomes ARAP, we can safely remove the line search.

The individual constituents of these matrices are defined as follows:

\begin{itemize}
\item $\boldsymbol{B}=\Blbs$ is the linear blend skinning Jacobian matrix for our secondary weights $\boldsymbol{W}$, as computed by \refeq{linear-blend-skinning-matmul}. 
\item  $\boldsymbol{G}_{9} \in \mathbb{R}^{9 \numclusters \times 9\numtets}$ is our exploded cluster grouping matrix, computable from our per-cluster labels $\boldsymbol{l}$ and our rest geometry, that performs a mass-weighed averaging of the 9 deformation gradient quantities for all tetrahedra belonging to a cluster. 

\item 
$\boldsymbol{K} \in \mathbb{R}^{9 \numtets \times 3n}$ is a standard vector gradient operator, which maps displacements of vertices to deformation gradients on tetrahedra  \cite{KimDynamicDeformables}.

\item 
$\boldsymbol{L} = \boldsymbol{K}^T \mathcal{U} \boldsymbol{V} \boldsymbol{K} \in \mathbb{R}^{3n \times 3n}$ is the heterogeneous Laplacian operator. ($\mathcal{U}$ and $\boldsymbol{V}$ are $9\numtets \times 9\numtets$ diagonal matrices containing the first Lam\'e parameter and volume respectively for each tetrahedron on their diagonal).

\item 
$\boldsymbol{M} \in \mathbb{R}^{3n \times 3n}$
is the vector mass matrix.

\item 
$\boldsymbol{J} \in \mathbb{R}^{3n \times \dimp}$ is the
rig Jacobian for our \textit{primary} rig.
\end{itemize}


We can leverage the fact that our subspace just performs linear blend skinning to project our low dimensional output $\boldsymbol{z}_{next}$ to the full space entirely on the GPU. We implement this projection step in the vertex shader, where we load the initial skinning weights as vertex attributes, and then pass the updated  $\boldsymbol{z}_{next}$ as a uniform in each draw call.


\section{Experiments \& Discussion}
We evaluate the effectiveness of our skinning subspace for deformation by comparing it to a suite of common deformation subspaces.  In our qualitative comparisons, we consider equal dimensional subspaces (e.g. 48 displacement modes = 12 degrees of freedom * 4 skinning eigenmodes). While this puts CPU computation on equal footing, we emphasize that our skinning eigenmodes use significantly less GPU memory in a vertex shader implementation ($12 \times$).

\subsection{Comparison to Modal Derivatives}
Modal derivatives \cite{BarbicJames:RealTimeSTVK} augment the primary displacement subspace with a second set of modes that aim to correct the primary modes as they fall out of date with large deformations. This effectively makes the subspace \emph{quadratic} instead of linear.
Unfortunately, modal derivatives are not closed under rotations. We show how this expresses itself in an elastic simulation in  \reffig{subspace-comparisons}. We compare 3 subspace simulations:  60 displacement modes, 11 displacement modes augmented with 49 modal derivatives, and 10 skinning modes (corresponding to 60 total d.o.f.s in 2D). While modal derivatives greatly enrich the space available to the elasto-dynamics, they provide unintuitively different deformations under rotations.
 By contrast, our subspace allows for the lowest energy at equilibrium, maintains a consistent equilibrium shape (and energy) no matter the orientation.

 \reffig{moray-rotation-equivariance} shows how this shortcoming manifests itself in our use case of fast complementary dynamics.  In this example, a user applies small repeated rotations about the y-axis on a single affine handle controlling the moray eel, mimicking a swiming motion. Rig momentum is allowed to leak from the rig to the eel to induce secondary motion. Then, the user rotated the eel about the z-axis 90 degrees, making the eel swim sideways.  As a user performs this rotation they would expect the resulting simulation to be identical up to a rotation as well, a property our 14 (168 d.o.f.s) skinning modes guarantee. Unfortunately the same cannot be said of a simulation using a modal derivative subspace of slightly greater size than ours.
 In this example, 17 primary modes augmented with a  full set of 153 modal derivatives (for a total of 170 d.o.f.s) were used to construct the modal derivative subspace. Like our skinning eigenmodes, was also constrained to satisfy the complementarity constraint by construction.
 
 Both methods benefit from performing a full-space projection step (to update the visualizion) on the GPU. As a linear blend skinning subspace, we can perform this step at a much smaller memory cost for the vertex shader. For a full space projection step with $k=d(d+1)m$ degrees of freedom, modal derivatives would require $k$-vec4's in vertex shader memory, whereas ours would only require $m$.

\begin{figure}
\includegraphics[width=\linewidth,keepaspectratio]{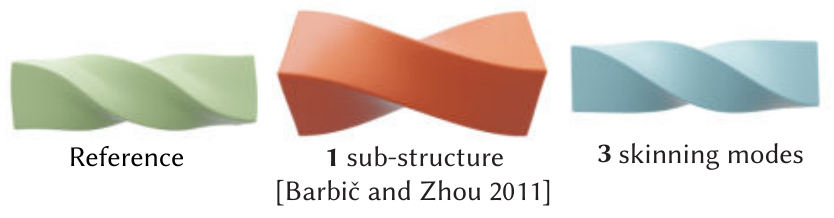}\timestamp{\tsBarModalWarpingComparison}
\caption{ Our skinning subspace (right) can easily represent twisting motions, something displacement modes struggle with, even with localized rotation fitting. 
 \label{fig:modal-fitting-rotation-clusters-comparison}}
\end{figure}

\subsection{Comparison to Sub-Structuring and Modal Warping}
Perhaps the most intuitive way ensuring our subspace is closed under rotations is by warping the subspace manually with a single best fit rotation. This corresponds to a simplified version of both sub-structuring and modal warping \cite{Barbic:2011:RealTimeLargeDefoSubstructuring, KimJamesMultiDomainStitching, ModalWarping}. While this indeed would fix the simulation rotation equivariance problem, it comes at the cost of not being able to represent rotational motion within each sub-structure, as shown didactically in \reffig{modal-fitting-rotation-clusters-comparison}.
Increasing the number of sub-structures  would mitigate this, but still requires the user to warp the subspace associated with each vertex every timestep, a full space operation. 
Accelerating this update at each using the Fast Sandwich Transform (FST) proposed by \cite{KimJamesMultiDomainStitching} is certainly possible for terms \emph{linear} in $\boldsymbol{B}$. However, it requires the reassembly of all linear pre-computed matrices involving $\boldsymbol{B}$, summing the contributions of each of the 9 rotation parameters \textit{for each} new sub-structure. Albeit accelerated compared to a na{\"\i}ve update, the FST update nevertheless becomes the bottleneck of our method, limiting the richness of the overall dynamics we can get for real-time rates. The FST also incurs a memory cost; in 3D, the aforementioned precomputed matrices must be stored separately 9 times \textit{for each} new rotation cluster.  On top of this, the FST would not allow us to efficientily update terms that are non-linear with respect to $\boldsymbol{B}$, such as our system matrix or our Cholesky prefactorization.

By contrast, our skinning subspace can represent twisting motions with a single static subspace computed once in a precomputation phase and \emph{never} updated again.

\begin{figure}
\includegraphics[width=\linewidth,keepaspectratio]{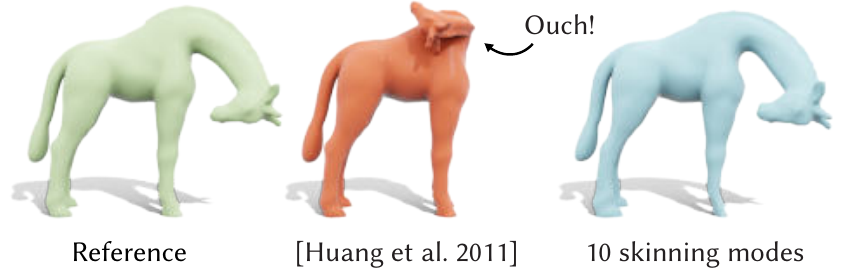}\timestamp[-0.125cm]{\tsGiraffeRSComparison}
\caption{ Rotation Strain 
  coordinates \cite{RScoords} are not well suited for reconstructing shapes undergoing rotational motion, as is commonly imposed by control rigs. Our subspace can fit this complex localized rotational motion with only 10 skinning modes.
 \label{fig:rs_comparison}} 
\end{figure}

\begin{figure}[!t]
\includegraphics[width=\linewidth,keepaspectratio]{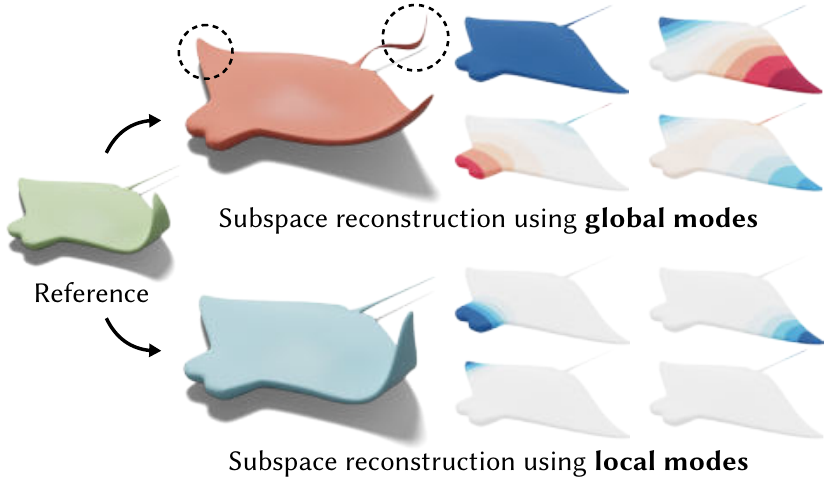}
\caption{Adopting  ICCM \cite{Brandt2017CompressedVibrationModesofElasticBodies} allows us to reconstruct the higly localized reference deformation (left) using a least squares fit, a task in which traditional global modes struggle. We build our local modes (bottom) with a sparsity regularization parameter $\eta=100$ and only fit the first 5 modes for each. 
\label{fig:stingray-local-vs-global-modes.pdf}}
\end{figure}

\subsection{Rotation Spanning vs. Rotation Equivariance} 
\edit{
\reffig{rotation-fitting} highlights the difference between rotation spanning and closure under rotations. The left side shows a least squares fitting task carried out in a subspace, attempting to fit a rotation of the rest geometry. If the subspace does not span rotations, like standard displacement modes, it will not be able to reconstruct the target rotated rest geometry. Adding rotational degrees of freedom\cite{Terzopoulos1988} by definition fix this problem and provides an equivalent solution to using a single constant skinning mode as our subspace. 
The right side of this figure does \emph{not} attempt to fit rotations, but attempts to fit a subspace displacement applied on top of a user-rotated frame. Specifically it shows that without closure under rotations, how well you reconstructing this displacement becomes a frame dependant process. Even if we added an extra set of affine degrees of freedom to our subspace (with error shown in purple), these would provide negligible help.  
}


\subsection{Comparison To Rotation Strain Coordinates}
Rotation strain coordinates  \cite{RScoords} fix linearized deformation artifacts that arise out of simulations using linear subspaces by projecting it to an energetically favorable pose.
They decompose the tetrahedron's deformation gradient 
$\boldsymbol{F} = \boldsymbol{r} + \boldsymbol{s} + \boldsymbol{I}$ 
into a symmetric shear part $\boldsymbol{s}$ and antisymmetric linearized rotation part $\boldsymbol{r}$.
The linearized rotation is then projected to a full rotation matrix via an exponential map 
$\boldsymbol{R} = \text{exp}(\boldsymbol{r})$, from which the deformation gradient is reconstructed as 
$\boldsymbol{F}' = \boldsymbol{R}(\boldsymbol{s} + \boldsymbol{I})$. 
The vertex positions $\boldsymbol{x}$ are determined by minimizing 
$|| \boldsymbol{K} \boldsymbol{x} - \boldsymbol{f}'||_{\boldsymbol{M}}$,
where $\boldsymbol{f}' = \text{vec}(\boldsymbol{F}')$.
As \reffig{rs_comparison} shows however, this formulation depends on the input deformation to be composed mainly of shear motions (as a normal displacement subspace would). If a user manipulating a shape through a rig created a rotation on the neck of the giraffe, Rotation Strain coordinates lead to an undesired reconstruction of the shape, 
rendering them less suitable for reconstruction over rig-like motion.
In fact, we can verify that for an input 2D rotational motion about the $z$-axis of $\frac{\pi}{2}$ leads to a target deformation gradient 
%
%
%
%
\begin{align}
\boldsymbol{F}' &=  \begin{bmatrix}
0 & 0 & 0 \\
0 & 0 & 0 \\
0 & 0 & 1
\end{bmatrix} \nonumber 
\end{align} 
which will act to collapse neighboring vertices to lie solely along the z-axis.
Furthermore, the Rotation Strain coordinates position fitting step may undo any constraints enforced in earlier steps, including our complementarity constraint. Identifying how to optimally enforce such constraints in the rotation strain coordinates pipelines remains an interesting avenue for future work.
\begin{figure}
\includegraphics[width=\linewidth,keepaspectratio]{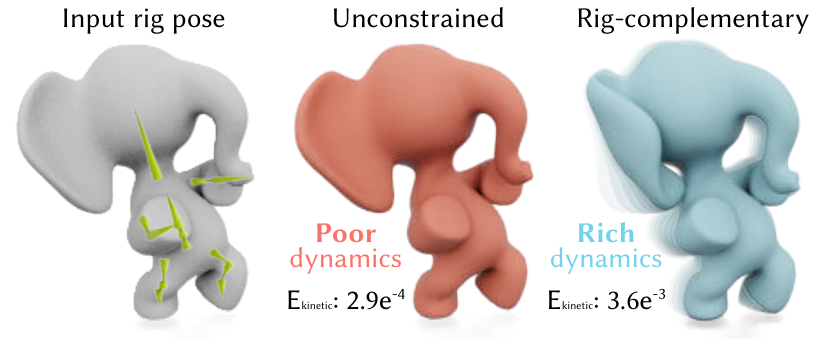}\timestamp{\tsConstrainedVsUnconstrained}
\caption{\label{fig:constrained_vs_unconstrained_reduced_cd_sim} Our rig-complementary subspace (right) composed of 14 skinning modes provide richer dynamics than an equal sized unconstrained subspace(left).}
\end{figure}

\subsection{Localized Skinning Subspaces}
Thanks to our generalized eigenvalue formulation, we can greatly benefit from prior work in localizing and sparsifying PCA/Modal analysis subspaces \cite{Nasikun2018, melzi2018localized, compressedModesADMM}. As an example, we can promote sparsity and locality in our modes to by applying the  \emph{iterated convexification for
compressed modes} (ICCM) \citep{Brandt2017CompressedVibrationModesofElasticBodies} algorithm. This amounts to simply adding an L1 regularization term to our minimization in \refeq{gevp-skinning-modes-constrained}. The result is a subspace that can faithfully capture local deformations as shown in
\reffig{stingray-local-vs-global-modes.pdf}. 
The skinning modes' closure under rotation, critical for producing rotation equivariant results, naturally
inherits the locality presented by the skinning weights:
observe that Eq.~\eqref{eq:linear-blend-skinning}
is a superposition of the closure of every mode $b$ in isolation: 
\begin{align}
w_{ib} \left( \boldsymbol{R}\boldsymbol{T}_b\right) \begin{bmatrix} \boldsymbol{x}_{0i} \\ \boldsymbol{1} \end{bmatrix} = \boldsymbol{R} \left(w_{ib} \boldsymbol{T}_b \begin{bmatrix} \boldsymbol{x}_{0i} \\ \boldsymbol{1} \end{bmatrix}\right)\, .
\end{align}

\newpage
\subsection{Difficulty Capturing High Frequency Motion}
 \begin{wrapfigure}[9]{r}{5cm}
\includegraphics[width=\linewidth,keepaspectratio]{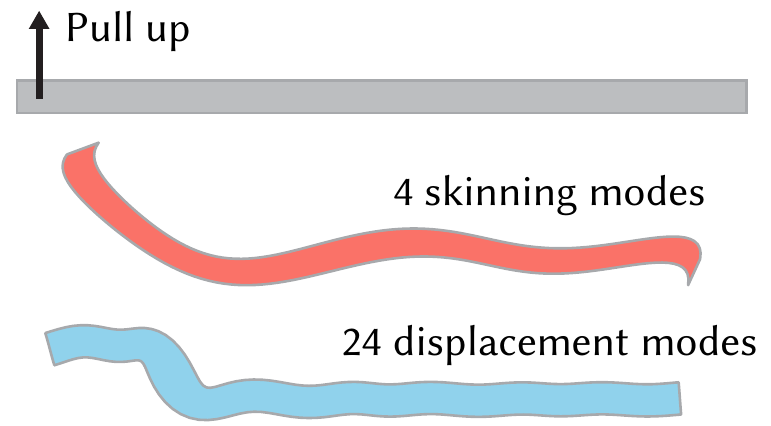}
\end{wrapfigure}

\edit{
Every secondary skinning weight in our subspace is associated with 12 affine degrees of freedom in 3D, as shown in \reffig{motions-producible-by-skinning-modes} (and 6 in 2D). While this is the key to many of the rotational qualities of our subspace, it also means that our subspace requires many degrees of freedom to represent certain high frequency motions. The inset shows a floppy bar being pulled at its left extremity, creating an elastic wave that ripples from left to right. A traditional displacement subspace can capture such high frequency motion, whereas our skinning subspace suffers from global artifacts with the same number of degrees of freedom.}


\subsection{Comparison to Coarsening Meshes}
A common approach to accelerating elasto-dynamics for real-time applications is to embed a fine mesh inside a coarse one, solve the elasto-dynamics on the coarse mesh, and then map the final motion back to the fine mesh via the embedding.

This corresponds to a very specific subspace $B_{embed}$; one that is very sparse, highly localized, can represent rotations, and maintains simulation rotation equivariance. 
Unfortunately, its construction from a coarse mesh embedding groups fine scale features that are close in Euclidean space whose motion should not be correlated. This leads to visible embedding artifacts as shown in \reffig{coarsening-meshes}. 
 \
To compare, our subspace couples motion of vertices as measured by the elastic energy.

\subsection{Discussion on Material Sensitivity}
An important feature that distinguishes our skinning weights from those of \citet{1Gilles2011, Wang:2015:LinearSubspaceDesign, Lan2020} is that our skinning weights are \emph{material}-aware. We show that this directly leads to richer dynamics when dealing with subspace simulations of heterogeneous materials, as shown in \reffig{heterogeneous-skinning-modes}.

\begin{figure}
\includegraphics[width=\linewidth,keepaspectratio]{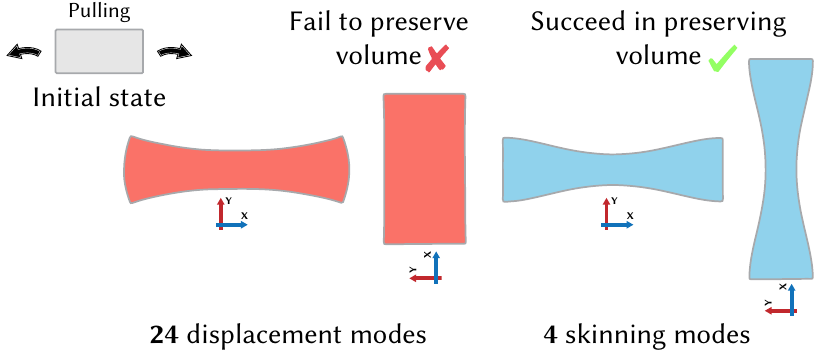}
\caption{ Our skinning subspace can easily capture volume-preserving effects for non-zero poisson ratio, and maintains such high quality deformations under any rotations of the rest frame.
 \label{fig:co-rotational-material-example-2D}}
\end{figure}
Because our final  weight space Hessian in \refeq{weight-space-hessian} does not capture any coupling interactions between the dimensions of our simulation it is invariant to changes in the Poisson ratio, which appear in the off-diagonal entries of the full space hessian $\boldsymbol{H}$. Fortunately, our skinning subspace can form non-uniform scales and shears, which allow it to excellently capture volume preserving effects with very few skinning modes, as shown in \reffig{co-rotational-material-example-2D}.

\subsection{Importance of Constraining Subspace}
\edit{One big difference between our skinning subspace and prior methods \cite{Faure2011, Tycowicz2013, Brandt2018HyperReducedPD} is that we can impose homogeneous equality constraints on our skinning weights by construction.}
Without this property, we'd have to impose our constraint at run-time. A user is burdened with having to select a subspace with more degrees of freedom than the constraint set in order to avoid an over-constrained system. To make things worse, the constraints for our primary rig can quickly climb in dimensionality; for a linear blend skinning control rig, every new bone brings with it $d(d+1)$ new columns in our complementarity constraint matrix $\boldsymbol{J}$.
\reffig{constrained_vs_unconstrained_reduced_cd_sim} shows a primary control rig composed of 14 bones (168 constraints). In order for a simulation to avoid being over-constrained, the user would then have to prescribe over 14 skinning modes (168) degrees of freedom. 
In contrast, our subspace frees the user from this dilemma, and allows the user to pick the dimensionality of the subspace without worrying about the well-posedness of the simulation.

\begin{wrapfigure}{l}{3.65cm}
\includegraphics[width=\linewidth,keepaspectratio]{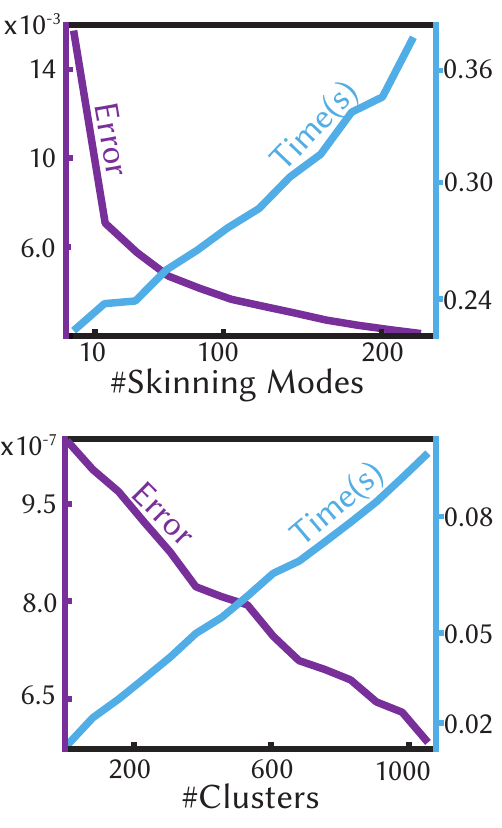}

\end{wrapfigure}
\subsection{Subspace Ablation}

\edit{The inset shows an ablation study on the number of modes (top) and the number of clusters (bottom). 
For both ablations, we compare the displacement incurred after a single timestep's displacement to a reference full space displacement, and compute the L2 error $|| \boldsymbol{u} - \boldsymbol{u}_{\mathrm{ref}} ||^2$. For the ablation on the modes, we fix the number of clusters to 3000. 
For the ablation on the clusters, we fix the number of skinning modes to 200. \newline \newline
}

\begin{figure}
\includegraphics[width=\linewidth,keepaspectratio]{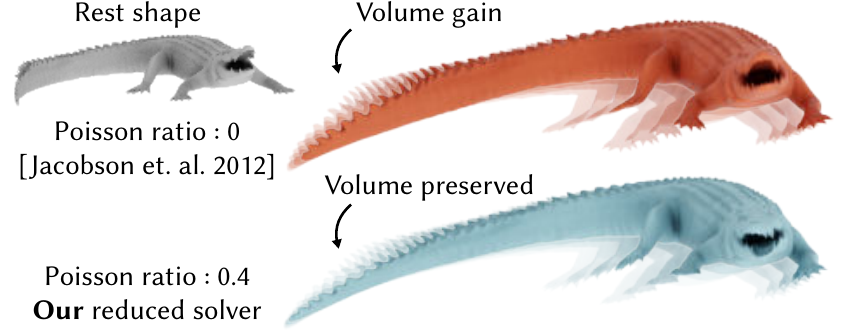}\timestamp{\tsCrocodileCorotVsARAP}
\caption{ We observe volume preserving effects made possible by accommodating co-rotational elasticity.  \label{fig:corotational-vs-arap-dragon-head} }
\end{figure}

\subsection{Evaluating Co-Rotational Elasticity}
We generalize the clustering based local global solver of \citet{Jacobson-12-FAST} to work with a more general material model of Linear Co-Rotational Elasticity \cite{McAdams2011}. We show a simple example in \reffig{co-rotational-material-example-2D} that shows how our subspace simulation with 50 clusters and 10 skinning modes can incorporate smooth volume preserving effects. 

 \reffig{corotational-vs-arap-dragon-head} highlights that these volume preserving effects have a noticeable visual impact on the secondary dynamics. 
 Unfortunately, such non-linear effects aren't free, a non-zero Poisson's ratio incurs the cost of tacking on a line search at the end of our solver's global step \cite{QuasiNewtonLiu2017}. Evaluating the energy for the line search requires re-computing per-cluster deformation gradients, a $O(rm)$ operation that corresponds to the slowest part in our simulation pipeline for large modes and large clusters. While this does sometimes hinder the speed of our simulation step, we have found that there are large portions of the mode/cluster parameter space that easily accommodate this computational overhead while still providing rich dynamics.

\begin{table}
\rowcolors{2}{white}{cyan!25}
\caption{Computing 10 of our skinning modes (120 d.o.f.s) is much faster than computing 120 displacement modes (120 d.o.f.s) subspace of an equivalent size.
\label{table:mode-computation-timings}}
\begin{tabular}{c|c|c|c}
      \textbf{Mesh} & \textbf{\#Vertices} &\textbf{ Disp.(s)} & \textbf{Ours(s)}  \\
     Elephant & 7842  & 0.782  &\textbf{ 0.135}\\
     Bulldog & 31368  & 6.76  & \textbf{0.830}\\
     XYZ Dragon & 99813  & 62.3 & \textbf{5.14} \\
     King Ghidora & 294033  & 143.4 & \textbf{14.07}  
\end{tabular}
\end{table}

 \begin{wrapfigure}{l}{5cm}
\includegraphics[width=\linewidth,keepaspectratio]{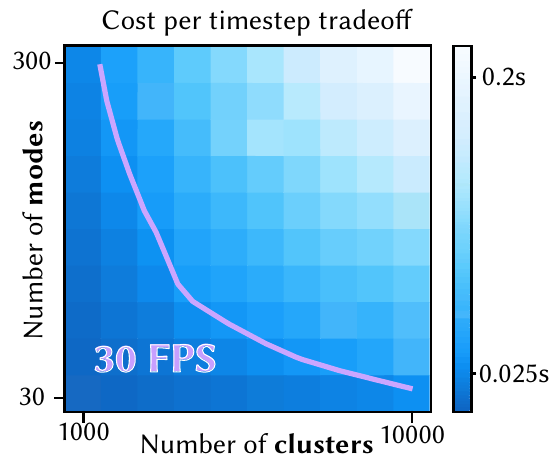}
\end{wrapfigure}
\subsection{Mode/Cluster Pareto-Front}
We evaluate how our per-timestep cost changes with the size of our subspace skinning modes and the number of clusters used. The inset highlights the 30 FPS real-time pareto front. For each trial, the same timestep was run 100 times until convergence on an ARAP elasticity model.

\subsection{Computing Our Modes}

Because our weight space Hessian is $n \times n$, compared to the traditional full space $n(d) \times n(d)$ Hessian, solving for our subspace is naturally faster for higher dimensions, as shown in Table \ref{table:mode-computation-timings}. Applying the complementarity constraint on our modes means augmenting the weight space Hessian $\boldsymbol{H}_w$ in our weight computation with our weight-space complementarity constraint $\boldsymbol{J}_w \in \mathbb{R}^{n \times \dimp (d)(d+1)}$. This makes mode computation naturally more computationally expensive for more complicated rigs.  The inset shows how changing the number of rig parameters in a control rig affects how long it takes to compute 10 skinning modes on the elephant mesh.

 \begin{wrapfigure}{r}{5cm}
\includegraphics[width=\linewidth,keepaspectratio]{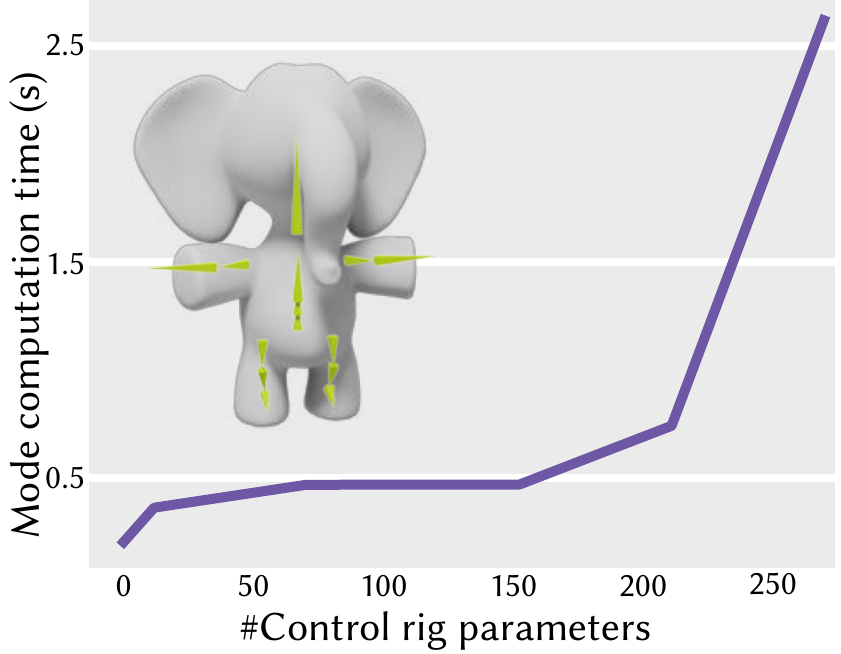}
\end{wrapfigure}

\section {Results}

\subsection{Rigid Body Augmentation}
\begin{figure}
\includegraphics[width=\linewidth,keepaspectratio]{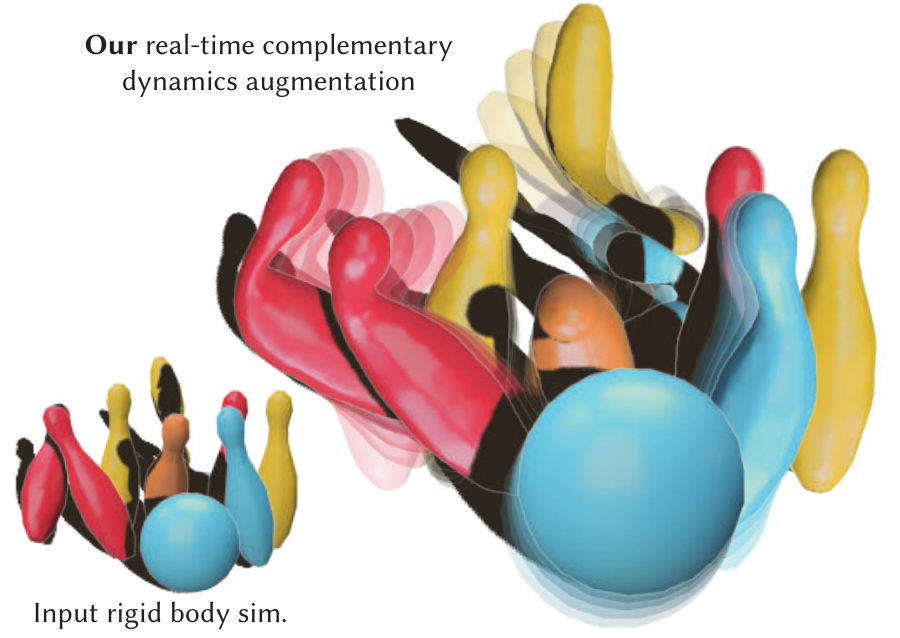}\timestamp{\tsRigidBodyGame}
\caption{\label{fig:rigid_body_augmentation} Augmenting a real-time rigid body simulation with complementary dynamics in real-time. }
\end{figure}

Rigid body simulations are key ingredients in many video games and virtual reality applications. A rigid body can also also be interpreted as a very simple specific case of a linear blend skinning, with only a single bone of constant weight equal to 1 across all the vertices.
We exploit this analogy to augment real-time rigid-body simulations with secondary effects (\reffig{rigid_body_augmentation} shows screenshots from an interactive bowling game, where the pins are stylized with secondary effects.)

\subsection{Augmenting Mixamo Characters with Secondary Motion}
\begin{figure}
\includegraphics[width=\linewidth,keepaspectratio]{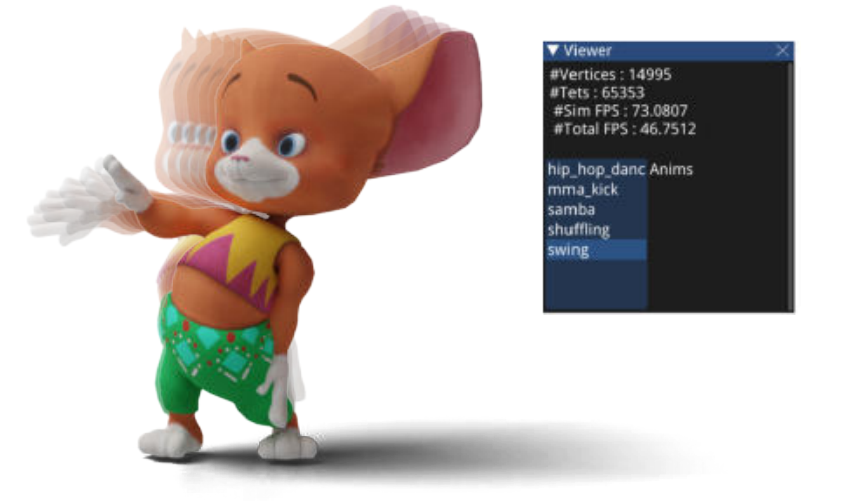}\timestamp{\tsMixamoRigSwitchingApp}
\caption{Augmenting and browsing various Mixamo character animations in real-time.
 \label{fig:mixamo-rig-switching-app} }
\end{figure}
Animators often design large suites of rig animations, often having to browse and edit many of these in rapid succession. Adding on to the tediousness of this process, they also have to foresee how these animations will look when augmented with secondary motion. 
Our method allows a user to browse through any number of rig-animations and immediately observe the secondary secondary dynamics in real-time. We demonstrate this for a character downloaded from the Mixamo~\shortcite{Mixamo} website in \reffig{mixamo-rig-switching-app}.

%

\subsection{Secondary Dynamics for Digital Avatars}
\begin{figure}[!ht]
\includegraphics[width=\linewidth,keepaspectratio]{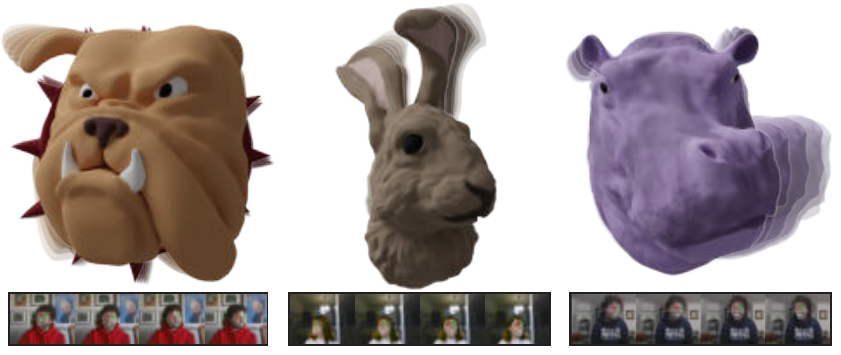}\timestamp[-0.125cm]{\tsMediaPipeFace}
\caption{ Augmenting mediapipe's face tracking with real-time secondary motion on digital avatars.
\label{fig:dynamic_snapchat_filters}}
\end{figure}

\begin{figure}[!ht]
\includegraphics[width=\linewidth,keepaspectratio]{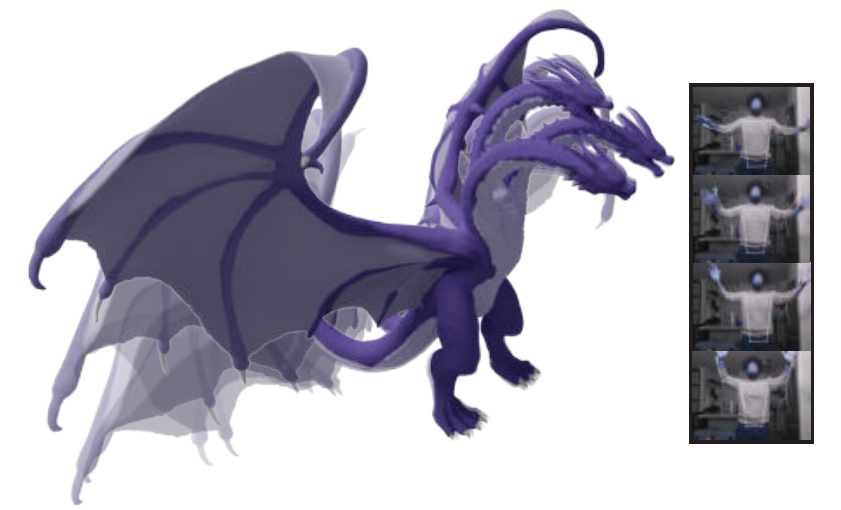}\timestamp{\tsMediaPipePose}
\caption{ Augmenting mediapipe's posetracking with real-time secondary motion on digital avatars.
\label{fig:pose_tracker_king_ghidorah}}
\end{figure}

Digital avatars are key to expressing personality and style in virtual environments. We show that our Fast Complementary Dynamics can be plugged into  very simple camera based pose trackers made available by mediapipe \cite{lugaresi2019mediapipe} to breath life into these digital avatars. \reffig{dynamic_snapchat_filters} demonstrates augmenting human faces tracked by mediapipe's Faces solution with secondary motion. 
Similarly, \reffig{pose_tracker_king_ghidorah} augment skeletons tracked by mediapipe's Pose solution with real-time secondary motion. In practice the bottleneck of this application became mediapipe's own Pose solution, requiring us to separate the tracker from the dynamics and place them into two separate threads.
\section{Limitations and Future Work}
Compared to the original complementary dynamics \cite{Zhang:CompDynamics:2020}, we assume that the user's rig is linear.
While non-linear rigs are less common in real-time settings, it would still be interesting to accommodate them (e.g., \cite{Kavan-08-SDQ}, perhaps via low-rank updates to the reduced system matrices in Alg.~\ref{alg:simulationStep} \cite{RSPHahn}. Our reduction techniques are in many ways agnostic to the choice of elastic potential. In future work, we would like to explore beyond co-rotational models. Our skinning eigenmodes easily facilitate rig-orthogonality constraints for complementary dynamics; their other good qualities suggest they could be useful beyond our target application, applied to more general elasticity problems (e.g., structural analysis or inverse design).

We leave comprehensive collision handling as future work.
While rigid-body augmentation provides a drop-in heuristic for collision effects for relatively stiff or fast moving objects, it would be interesting to explore more accurate methods, perhaps combining our contributions with cubature techniques \cite{HarmonSubspaceLocalDeformation2013}.
For real-time VR avatar, applications we found the bottleneck for quality lies in the tracking and mapping of user movements to primary rig controls. Our secondary effects would immediately inherit any improvements in these active research areas.

We derive skinning eigenmodes by considering distributions of translations. This sidesteps the issue of determining an origin and scale associated with each mode.  It would be interesting to treat these as variables (to achieve theoretical optimality over distributions of general affine transformations). This appears non-trivial and irreducible to a generalized eigenvalue problem

We have presented a novel subspace for deformation that is well suited for augmenting real-time rig animations with secondary, complementary motion.
Our computation is well-balanced between fast small iterations on the CPU and memory-efficient, standard-pipeline vertex shaders on the GPU.
In future work, we are interested in not just adding features to our elasticity effects, but also considering complementary dynamics more broadly into domains such as fluid simulation, electrodynamics, and crowds.
We hope that our work also serves as general recipe for translating 
complementary dynamics to real-time scenarios.

\begin{acks}
This project is funded in part by NSERC Discovery (RGPIN2017–05235,
RGPIN-2021-03733,
RGPAS–2017–507938) New Frontiers of Research Fund (NFRFE–201),
the Ontario Early Research Award program, the Canada Research
Chairs Program, a Sloan Research Fellowship, the DSI Catalyst Grant program and gifts by Adobe Inc. Otman Benchekroun was funded by an NSERC CGS-M scholarship, while Jiayi Eris Zhang is funded by a Stanford Graduate Fellowship.
We thank David I.W. Levin, Danny M. Kaufman, Doug L. James, Ty Trusty, Sarah Kushner and Silvia Sell{\'{a}}n for insightful conversations. We thank Alejandra Baptista Aguilar for early discussion and adoption in 3D environments, and for providing some of the models in the paper. We thank Silvia Sell{\'{a}}n, Selena Ling, Shukui Chen, Aravind Ramakrishnan and Kinjal Parikh for proofreading.  We thank Xuan Dam and John Hancock for technical and administrative assistance throughout the course of this project.

\end{acks}

\bibliographystyle{ACM-Reference-Format}
\bibliography{sample}

\appendix

\section{Proofs}
\subsection{A subspace simulation is rotation equivariant if and only if the subspace is closed under rotations} 
\label{sec:appendix-proof-sim-rotation-equivariance}
Consider the minimization of a rotation invariant elastodynamic energy $E(\boldsymbol{u} + \boldsymbol{x})$, where $\boldsymbol{u}, \boldsymbol{x} \in \mathbb{R}^{3n}$. By rotation equivariance we mean that 
\begin{align}
 \forall \boldsymbol{R}\, , \ \  
  \forall \boldsymbol{x}\, , \ \
    \argmin_{\boldsymbol{u}} E(\boldsymbol{u}+  \repR\boldsymbol{x} ) = \repR \argmin_{\boldsymbol{v}} E(\boldsymbol{v} +  \boldsymbol{x} ) \ . \nonumber
\end{align}
Here and henceforth the domains for $\boldsymbol{R}$ and $\boldsymbol{x}$ shall be the rotations $\mathcal{SO}(3)$ and the positions $\mathbb{R}^{3n}$, respectively.



In a full space simulation, the above holds. For a subspace simulation where $\boldsymbol{u} = \boldsymbol{B} \boldsymbol{z}$, and  $\mathrm{Col}(\boldsymbol{B}) \subset \mathbb{R}^{3n}$, the equivalent statement,
\begin{align}
 \forall \boldsymbol{R}\, , \ \  
  \forall \boldsymbol{x}\, , \ \
    \boldsymbol{B} \argmin_{\boldsymbol{z}} E(\boldsymbol{Bz}+  \repR\boldsymbol{x} ) = \repR \boldsymbol{B} \argmin_{\boldsymbol{w}} E(\boldsymbol{Bw} +  \boldsymbol{x} ) \, ,
    \label{appendix-eq:rotation-equivariant-subspace-sim}
\end{align}
no longer holds for all subspaces. We shall prove that the above holds \emph{if and only if} our subspace is closed under rotations, 
\begin{align}
    \forall \, \boldsymbol{R}\in SO(3), \  \forall \, \boldsymbol{z}\in\mathbb{R}^m, \  \exists \, \boldsymbol{w}\in\mathbb{R}^m, \,  \boldsymbol{Bz} = \repR \boldsymbol{B} \boldsymbol{w}  \ .\quad 
\label{appenix-eq:closed-under-rotations}
\end{align}

\subsubsection{Forward Proof}
Given that our subspace satisfied closure under rotations (Eq. \eqref{appenix-eq:closed-under-rotations}), we prove that Eq. \eqref{appendix-eq:rotation-equivariant-subspace-sim} holds. 

\paragraph{Theorem: 
$\eqref{appenix-eq:closed-under-rotations} \rightarrow \eqref{appendix-eq:rotation-equivariant-subspace-sim} $} \textbf{Proof:}
Consider (without loss of generality) some rotation $\boldsymbol{R}$ and position $\boldsymbol{x}$.
Let $\boldsymbol{z}^*$ be the minimizer of the left hand side of Eq. \eqref{appendix-eq:rotation-equivariant-subspace-sim}:
\begin{align}
\boldsymbol{z}^* = \argmin_{\boldsymbol{z}} E(\boldsymbol{Bz}+  \repR\boldsymbol{x} ) \label{appendix-eq:definition-z-star}
\end{align}
By closure of our subspace under rotations, 
$ \exists \, \boldsymbol{w}^* \, , \  \boldsymbol{Bz^*} = \repR\boldsymbol{Bw^*}. $

The minimum energy is thus
\begin{align} 
E^* = E(\boldsymbol{Bz}^* + \repR \boldsymbol{x}) = E(\repR\boldsymbol{B} \boldsymbol{w}^*  + \repR \boldsymbol{x}) 
=E(\boldsymbol{B} \boldsymbol{w}^* +  \boldsymbol{x})
\nonumber 
\end{align}
where (from left to right) the equalities use that $\boldsymbol{z}^*$ is a minimizer, $\boldsymbol{B}$ is closed under rotations, and $E(\cdot)$ is rotation invariant, respectively. Consequently, $\boldsymbol{w}^* = \argmin_{\boldsymbol{w}} E(\boldsymbol{Bw}+  \boldsymbol{x} )$, which we substitute along with \eqref{appendix-eq:definition-z-star} into \eqref{appenix-eq:closed-under-rotations} to complete the proof.$\qed$

\subsubsection{Backward Proof}
Given our subspace simulation is rotation equivariant (Eq. \eqref{appendix-eq:rotation-equivariant-subspace-sim}), we prove that Eq. \eqref{appenix-eq:closed-under-rotations} holds.

\paragraph{Lemma} 
\begin{align}
\forall \boldsymbol{R}^* \in \mathcal{SO}(3),  \forall \boldsymbol{z}^* \in \mathbb{R}^{m},\\
\exists \boldsymbol{x}^*\, , \ \ \argmin_{\boldsymbol{z}} E(\boldsymbol{Bz} + \rep{\R^*}\boldsymbol{x}^*) = \boldsymbol{z}^* \, . \end{align}

\textbf{Proof:} 
Let $\boldsymbol{\tilde{z}}$ be some minimizer of $E(\boldsymbol{Bz})$ by construction. 
To satisfy the Lemma we choose
$\boldsymbol{x}^* = \rep{\R^*}^{-1} \boldsymbol{B} (\boldsymbol{\tilde{z}} - \boldsymbol{z}^*)$, because
\begin{align}
\argmin_{\boldsymbol{z}} E(\boldsymbol{Bz} + \rep{\R^*} \boldsymbol{x}^*) \nonumber  = \\
\argmin_{\boldsymbol{z}} E(\boldsymbol{Bz} + \rep{\R^*}\rep{\R^*}^{-1}\boldsymbol{B} (\boldsymbol{\tilde{z}} - \boldsymbol{z}^*))  = \\
\argmin_{\boldsymbol{z}} E(\boldsymbol{B} (\boldsymbol{z} + \boldsymbol{\tilde{z}} - \boldsymbol{z}^*)) \, .
\nonumber 
\end{align}
Thus $w = \boldsymbol{z} + \boldsymbol{\tilde{z}} - \boldsymbol{z}^*$ minimizes $E(\boldsymbol{Bw})$, and so does $w = \boldsymbol{\tilde{z}}$ (by construction). Since $w=w$,  then $\boldsymbol{z} = \boldsymbol{z}^*$ minimizes $E(\boldsymbol{B}\boldsymbol{z} + \rep{\R^*} \boldsymbol{x}^* )  \qed$.

\paragraph{Theorem: 
$\eqref{appendix-eq:rotation-equivariant-subspace-sim} \rightarrow \eqref{appenix-eq:closed-under-rotations}$} \textbf{Proof:}
For any given $\boldsymbol{R}^*$, $\boldsymbol{z}^*$, we produce $\boldsymbol{w}^*$ satisfying
\begin{align}
\boldsymbol{B} \boldsymbol{z}^* = \rep{\R^*} \boldsymbol{B} \boldsymbol{w}^*  \, . \nonumber
\end{align}

Pick a specific $\boldsymbol{R}^*, \boldsymbol{z}^*$. Using the Lemma, 
we pick $\boldsymbol{x}^*$ such that $\boldsymbol{z}^*$ 
minimizes $E(\boldsymbol{Bz} + \rep{\R^*}\boldsymbol{x}^*)$.
By the rotation equivariance of our minimization,
\begin{align}
\boldsymbol{B} \argmin_{\boldsymbol{z}} E(\boldsymbol{B} \boldsymbol{z}  + 
\rep{\R^*} \boldsymbol{x}^*) &=  \rep{\R^*} \boldsymbol{B} \argmin_{\boldsymbol{w}} E(\boldsymbol{B} \boldsymbol{w}  + \boldsymbol{x}^* ) \nonumber \\
\boldsymbol{B} \boldsymbol{z}^*  &=  \rep{\R^*} \boldsymbol{B} \argmin_{\boldsymbol{w}} E(\boldsymbol{B} \boldsymbol{w}  + 
\boldsymbol{x}^*)  \nonumber \\
&=  \rep{\R^*} \boldsymbol{B}\boldsymbol{w}^* \, ,
\nonumber 
\end{align}
where we have chosen
$\boldsymbol{w}^*  = \argmin_{\boldsymbol{w}} E (\boldsymbol{B} \boldsymbol{w} + \boldsymbol{x}^*) \, .\qed
$

\section{Matrix Derivations}
\subsection{Weight Space Skinning Jacobian Products}
\label{appendix-sec:weight-space-skinning-jacobians}
The product $\boldsymbol{J}^T\boldsymbol{B}_{lbs}$ frequently arises in our derivations. Here $\boldsymbol{J} \in \mathbb{R}^{n(d) \times c}$ can be a generic matrix and $\boldsymbol{B}_{lbs} \in \mathbb{R}^{n(d) \times d(d+1)(m)}$ is our Linear Blend Skinning Jacobian matrix, which depends linearly on some set of weights $\boldsymbol{W} \in \mathbb{R}^{n(d) \times m}$. We leverage the relationship between $\boldsymbol{B}_{lbs}$ and $\boldsymbol{W}$ to rewrite the entries of the matrix product $\boldsymbol{J}^T \boldsymbol{B}_{lbs}$ entirely in terms of the weights:
\begin{align}
      \boldsymbol{B}_{lbs} = (\boldsymbol{I_d} \otimes \underbrace{((\boldsymbol{1}_m^T \otimes [\boldsymbol{V} \boldsymbol{1}_n])   \odot  (\boldsymbol{W} \otimes \boldsymbol{1}_{d+1}^T))}_{\boldsymbol{\mathcal{B}}}) \nonumber
\end{align}
 We assume the 3D case $d=3$, and restate the final result 2D. The full product whose entries we wish to relate linearly to our weights $\boldsymbol{W}$ is
\begin{align}
    \boldsymbol{J}^T (\boldsymbol{I_3} \otimes \mathcal{B}) 
\end{align}
Separating the mutiplying matrix into its 3 sets of columns:
\begin{align}
\boldsymbol{J}^T  = [\boldsymbol{J}_x^T  \; | \; \boldsymbol{J}_y^T  \; | \; \boldsymbol{J}_z^T ]
\end{align}
Expanding the rows of this product:
\begin{align}
    \begin{bmatrix}\boldsymbol{J}^T_x  \mathcal{B} \; |
    \; \boldsymbol{J}^T_y \mathcal{B} \; | 
    \; \boldsymbol{J}^T_z \mathcal{B} 
    \end{bmatrix} 
\end{align}
We focus on one set of columns $i \in \{x, y, z\}$ at a time:
\begin{align}
\boldsymbol{J}^T_i \mathcal{B} =  \boldsymbol{J}^T_i ((\boldsymbol{1}_m^T \otimes [\boldsymbol{X} \boldsymbol{1}_n])   \odot  (\boldsymbol{W} \otimes \boldsymbol{1}_{4}^T)))  \in \mathbb{R}^{j \times (4)m} \nonumber
\\
  \boldsymbol{J}^T_i ((\boldsymbol{1}_m^T \otimes [\boldsymbol{\bar{x}} \,\boldsymbol{\bar{y}} \, \boldsymbol{\bar{z}} \, \boldsymbol{1}_n])   \odot  ([\boldsymbol{w}_1 \boldsymbol{w}_2 ... \boldsymbol{w}_m] \otimes \boldsymbol{1}_{4}^T))) \nonumber
 \end{align}

 We recall the definition of the Linear Blend Skinning Jacobian
  where we split 
  
  $\boldsymbol{X} = [\boldsymbol{\bar{x}} \,\boldsymbol{\bar{y}} \, \boldsymbol{\bar{z}} ]$ into its individual columns.
 Expanding the two Kronecker products:
 \begin{align}
 \boldsymbol{J}^T_i ( [\underbrace{\boldsymbol{\bar{x}} \,\boldsymbol{\bar{y}} \, \boldsymbol{\bar{z}} \, \boldsymbol{1}_n}_{1} | ... | \underbrace{\boldsymbol{\bar{x}} \,\boldsymbol{\bar{y}} \, \boldsymbol{\bar{z}} \, \boldsymbol{1}_n}_b ]   \odot  [\underbrace{\boldsymbol{w}_1 \boldsymbol{w}_1 \boldsymbol{w}_1 \boldsymbol{w}_1}_1  | ... | \underbrace{\boldsymbol{w}_m \boldsymbol{w}_m \boldsymbol{w}_m \boldsymbol{w}_m}_{m}] )  \nonumber
 \end{align}
Distributing the products further:
 \begin{align}
  \boldsymbol{J}^T_i [\underbrace{([\boldsymbol{\bar{x}} \,\boldsymbol{\bar{y}} \, \boldsymbol{\bar{z}} \, \boldsymbol{1}_n] \odot  [\boldsymbol{w}_1 \boldsymbol{w}_1 \boldsymbol{w}_1 \boldsymbol{w}_1]}_1| ... | \underbrace{[\boldsymbol{\bar{x}} \,\boldsymbol{\bar{y}} \, \boldsymbol{\bar{z}} \, \boldsymbol{1}_n] \odot  [\boldsymbol{w}_m \boldsymbol{w}_m \boldsymbol{w}_m \boldsymbol{w}_m]}_m]  )  \nonumber
 \end{align}
Moving the matrix product inside, and looking at one block $q \in [1, ... m]$ at a time:
 \begin{align}
 & \boldsymbol{J}^T_i ( [ \boldsymbol{\bar{x}} \,\boldsymbol{\bar{y}} \, \boldsymbol{\bar{z}} \, \boldsymbol{1}_n] \odot  [\boldsymbol{w}_q \boldsymbol{w}_q \boldsymbol{w}_q \boldsymbol{w}_q])  \nonumber \\
 &= \boldsymbol{J}^T_d ([\boldsymbol{\bar{x}} \odot\boldsymbol{w}_q |  \boldsymbol{\bar{y}} \odot \boldsymbol{w}_q |\boldsymbol{\bar{z}} \odot \boldsymbol{w}_q | \boldsymbol{1}_n \odot \boldsymbol{w}_q]) \nonumber  
 \end{align}
 Rewriting the individual Hadamard products as diagonal matrix multiplication where we introduce the capitalized overbar notation $\boldsymbol{\bar{X}}, \boldsymbol{\bar{Y}}, \boldsymbol{\bar{Z}}$ to indicate the diagonal matrices whose diagonal entries are $\boldsymbol{\bar{x}},\boldsymbol{\bar{y}}, \boldsymbol{\bar{z}} $ respectively.
 \begin{align}
 &= \boldsymbol{J}^T_d ([\boldsymbol{\bar{X}}\boldsymbol{w}_q |  \boldsymbol{\bar{Y}}\boldsymbol{w}_q | \boldsymbol{\bar{Z}} \boldsymbol{w}_q |  \boldsymbol{w}_q])  \nonumber \\
  &= [ \boldsymbol{J}^T_d \boldsymbol{\bar{X}}\boldsymbol{w}_q |   \boldsymbol{J}^T_d \boldsymbol{\bar{Y}}\boldsymbol{w}_q |  \boldsymbol{J}^T_d \boldsymbol{\bar{Z}} \boldsymbol{w}_q | \boldsymbol{J}^T_d  \boldsymbol{w}_q]  \nonumber
 \end{align}
Applying this for all weights $b$:
\begin{align}
[\boldsymbol{J}^T_i \boldsymbol{\bar{X}} \boldsymbol{W} |   \boldsymbol{J}^T_i \boldsymbol{\bar{Y}} \boldsymbol{W} |  \boldsymbol{J}^T_i \boldsymbol{\bar{Z}} \boldsymbol{W} | \boldsymbol{J}^T_i  \boldsymbol{W}]  \nonumber
 \end{align}
 Finally, applying this to all 3 dimensions of $i$:
 \begin{align}
 \boldsymbol{J}^T \boldsymbol{B}_{lbs} = 
[& \boldsymbol{J}^T_x \boldsymbol{\bar{X}} \boldsymbol{W} \; | \;   \boldsymbol{J}^T_x \boldsymbol{\bar{Y}} \boldsymbol{W} \; | \;  \boldsymbol{J}^T_x \boldsymbol{\bar{Z}} \boldsymbol{W} \; | \; \boldsymbol{J}^T_x  \boldsymbol{W}] \; |\;  ... \nonumber  \\
& \boldsymbol{J}^T_y \boldsymbol{\bar{X}} \boldsymbol{W} \; |  \;  \boldsymbol{J}^T_y \boldsymbol{\bar{Y}} \boldsymbol{W} \; | \;  \boldsymbol{J}^T_y \boldsymbol{\bar{Z}}  \boldsymbol{W} \; | \; \boldsymbol{J}^T_y  \boldsymbol{W}] \; |\;  ... \nonumber
\\
& \boldsymbol{J}^T_z \boldsymbol{\bar{X}}  \boldsymbol{W} \; |\;    \boldsymbol{J}^T_z \boldsymbol{\bar{Y}} \boldsymbol{W} \; |\;   \boldsymbol{J}^T_z \boldsymbol{\bar{Z}}  \boldsymbol{W} \; | \; \boldsymbol{J}^T_z  \boldsymbol{W}] 
 \end{align}

We can derive our weight Skinning Jacobians by rewriting the entries  $\boldsymbol{B}_{lbs}$ alone in terms of the weights $\boldsymbol{W}$ as a special case of the above result.
If we choose $\boldsymbol{J}$ to be the identity, and slice out the 3 sets of columns belonging to the identity, we obtain our dimensional selection matrices $\boldsymbol{J} = \boldsymbol{I}_{3n} = [\boldsymbol{P}_x \boldsymbol{P}_y  \boldsymbol{P}_z]$. We can then rewrite :

 \begin{align}
 \boldsymbol{B}_{lbs} = 
[& \boldsymbol{P}_x \boldsymbol{\bar{X}} \boldsymbol{W} \; | \;   \boldsymbol{P}_x \boldsymbol{\bar{Y}} \boldsymbol{W} \; | \;  \boldsymbol{P}_x \boldsymbol{\bar{Z}} \boldsymbol{W} \; | \; \boldsymbol{P}_x  \boldsymbol{W}] \; |\;  ... \nonumber  \\
& \boldsymbol{P}_y \boldsymbol{\bar{X}} \boldsymbol{W} \; |  \;  \boldsymbol{P}_y \boldsymbol{\bar{Y}} \boldsymbol{W} \; | \;  \boldsymbol{P}_y \boldsymbol{\bar{Z}}  \boldsymbol{W} \; | \; \boldsymbol{P}_y  \boldsymbol{W}] \; |\;  ... \nonumber
\\
& \boldsymbol{P}_z \boldsymbol{\bar{X}}  \boldsymbol{W} \; |\;    \boldsymbol{P}_z \boldsymbol{\bar{Y}} \boldsymbol{W} \; |\;   \boldsymbol{P}_z \boldsymbol{\bar{Z}}  \boldsymbol{W} \; | \; \boldsymbol{P}_z  \boldsymbol{W}] 
 \end{align}
Where we identify  12 weight-Space Skinning Jacobian matrices:
\begin{align}
\begin{matrix}
\boldsymbol{A}_{1, 1} = \boldsymbol{P_x} \boldsymbol{\bar{X}}  &\boldsymbol{A}_{1, 2} = \boldsymbol{P_x} \boldsymbol{\bar{Y}} &\boldsymbol{A}_{1, 3} = \boldsymbol{P_x} \boldsymbol{\bar{Z}} &
\boldsymbol{A}_{1, 4} = \boldsymbol{P_x} \\
\boldsymbol{A}_{2, 1} = \boldsymbol{P_y} \boldsymbol{\bar{X}} &\boldsymbol{A}_{2, 2} = \boldsymbol{P_y} \boldsymbol{\bar{Y}} &\boldsymbol{A}_{2, 3} = \boldsymbol{P_y} \boldsymbol{\bar{Z}} &
\boldsymbol{A}_{2, 4} = \boldsymbol{P_y} \\
\boldsymbol{A}_{3, 1} = \boldsymbol{P_z} \boldsymbol{\bar{X}}  &\boldsymbol{A}_{3, 2} = \boldsymbol{P_z} \boldsymbol{\bar{Y}} &\boldsymbol{A}_{3, 3} = \boldsymbol{P_z} \boldsymbol{\bar{Z}} &
\boldsymbol{A}_{3, 4} = \boldsymbol{P_z}
\end{matrix}
\label{eq:weight-space-skinning-jacobian-3D}
\end{align}
We can derive the 2D case of these matrices by following the equivalent steps as above, but with one less set of columns for $\boldsymbol{I}$ and one less column in our rest positions $\boldsymbol{X}$:
\begin{align}
\begin{matrix}
\boldsymbol{A}_{1, 1} = \boldsymbol{P_x} \boldsymbol{\bar{X}}  &\boldsymbol{A}_{1, 2} = \boldsymbol{P_x} \boldsymbol{\bar{Y}} &\boldsymbol{A}_{1, 3} = \boldsymbol{P_x}  \\
\boldsymbol{A}_{2, 1} = \boldsymbol{P_y} \boldsymbol{\bar{X}} &\boldsymbol{A}_{2, 2} = \boldsymbol{P_y} \boldsymbol{\bar{Y}} &\boldsymbol{A}_{2, 3} = \boldsymbol{P_y}  
\end{matrix}
\label{eq:weight-space-skinning-jacobian-2D}
\end{align}

\section{Hyper-Reduced Local Global Solvers for Different Elastic Energies}
\subsection{Hyper-Reduced ARAP}
\label{sec:appendix-hyper-reduced-arap-elasticity}
The full space ARAP elastic energy can efficiently be written as:
\begin{align}
    E_{ARAP}(\boldsymbol{x}) = \sum_t^{k} m_t  \mu_t || \boldsymbol{F}_t -  \boldsymbol{R}_t||^2_F,
\end{align}
where $\boldsymbol{F}_t$ is our per-tet deformation gradient (a linear function of our degrees of freedom) and $\boldsymbol{R}_t$ is a best-fit rotation on the deformation gradient, a non-linear function of our degrees of freedom.
Expanding the square Frobemius norm, we can identify a quadratic component and non-linear component to our energy:
\begin{align}
    & E_{ARAP}(\boldsymbol{x}) = \underbrace{\sum_t^{k} m_t \mu_t tr(\boldsymbol{F}_t^T\boldsymbol{F}_t)}_{\Psi}  + \underbrace{ \sum_t^{k}- 2 m_t \mu_t tr(\boldsymbol{F}^T_t \boldsymbol{R}_t)}_{\Phi} \nonumber \\
     &=  \Psi(\boldsymbol{u})  + \Phi(\boldsymbol{u}). \nonumber 
\end{align}
We can rewrite this energy in its hyper-reduced form by recalling $\boldsymbol{x} = \boldsymbol{Bz}$ and approximating $\Phi \approx \tilde{\Phi}$. We expand our clustered non-linear component to the elastic energy:
\begin{align}
    E_{ARAP}(\boldsymbol{z}) &=   \Psi(\boldsymbol{z})  + \underbrace{\sum_c^{r}   - 2 m_c \mu_c tr(\boldsymbol{F}^T_c \boldsymbol{R}_c)  }_{\tilde{\Phi}} \nonumber \\ 
     &=  \Psi(\boldsymbol{z})  + \tilde{\Phi}(\boldsymbol{z} ). 
\end{align}
We then expose the source of the non-linearities, the best fit per-cluster rotation matrices, as auxillary degrees of freedom.
\begin{align}
    E_{ARAP}(\boldsymbol{z}, \boldsymbol{\tilde{R}}) &=    \Psi(\boldsymbol{z})  + \tilde{\Phi}(\boldsymbol{z}, \boldsymbol{\tilde{R}} ).
\end{align}
The hyper reduced local step becomes:
\begin{align}
    \boldsymbol{\tilde{R}} &=  \argmin_{\boldsymbol{\tilde{R}}}   \tilde{\Phi}(\boldsymbol{z}, \boldsymbol{\tilde{R}} ) \nonumber \\ 
    &= \argmin_{\boldsymbol{\tilde{R}}} \sum_c^{k} m_c  \mu_c tr(  \boldsymbol{F}^T_c \boldsymbol{R}_c)  \nonumber \\
    &= \text{PolarSVD}(\boldsymbol{\tilde{F}}) \nonumber
\end{align}
which is found through polar decomposition of the per-cluster deformation gradient $\boldsymbol{\tilde{F}}$, obtainable via a reshape operation of $\boldsymbol{\tilde{f}} = \boldsymbol{x}^T \boldsymbol{K}^T \boldsymbol{G} $.
The hyper reduced global step minimizes
\begin{align}
      \boldsymbol{z} &= \argmin_{\boldsymbol{z}} \Psi(\boldsymbol{z})   + \tilde{\Phi}(\boldsymbol{z}, \boldsymbol{\tilde{R}} ) \nonumber \\
       &= \argmin_{\boldsymbol{z}} \boldsymbol{z}^T \boldsymbol{B}^T \boldsymbol{K}^T \mathcal{U} \boldsymbol{V} \boldsymbol{K} \boldsymbol{B} \boldsymbol{z} +  \tilde{\Phi}(\boldsymbol{z}, \boldsymbol{\tilde{R}} ) \nonumber \\
         &= \argmin_{\boldsymbol{z}} \boldsymbol{z}^T \boldsymbol{B}^T \boldsymbol{L} \boldsymbol{B} \boldsymbol{z} + \tilde{\Phi}(\boldsymbol{z}, \boldsymbol{\tilde{R}} ), \nonumber 
\end{align}
where we introduce the heterogeneous Laplacian matrix $\boldsymbol{L} = \boldsymbol{K}^T \mathcal{U} \boldsymbol{V} \boldsymbol{K}$ for brevity. Here, $\boldsymbol{K} \in \mathbb{R}^{9k \times 3n} $ is the vector gradient operator over our mesh, mapping deformed coordinates to per-tet deformation gradients, $\boldsymbol{V} \in \mathbb{R}^{9k \times 9k}$ is a diagonal matrix of tetrahedron volumes, $\mathcal{U}  = (\boldsymbol{I}_9  \otimes \text{diag}(\boldsymbol{\mu}))) \in \mathbb{R}^{9k \times 9k}$ is a diagonal containing first lamé parameter $\mu_t$ for each tet $t$ along its diagonal. 

For ARAP, the global-step optimization is \emph{exactly} quadratic in the degrees of freedom $\boldsymbol{z}$ (because $\tilde{\Phi}(\boldsymbol{z}, \boldsymbol{\tilde{R}})$ is linear in $\boldsymbol{z})$, and can directly be found every local-global iteration via a single prefactorizable system solve:
\begin{align}
        \boldsymbol{0} &=  2 \boldsymbol{B}^T \boldsymbol{L} \boldsymbol{B} \boldsymbol{z} + 
        \frac{\partial \boldsymbol{x}}{\partial \boldsymbol{z}} \frac{\partial \boldsymbol{f}}{\partial \boldsymbol{x}}
        \frac{\partial \boldsymbol{\tilde{f}}}{\partial \boldsymbol{f}} 
        \frac{\partial \tilde{\Phi}}{\partial \boldsymbol{\tilde{f}}} \nonumber  \\
        &=  \boldsymbol{B}^T \boldsymbol{L} \boldsymbol{B} \boldsymbol{z}  +
        \frac{1}{2} \boldsymbol{B}^T \boldsymbol{K}^T \boldsymbol{G}
        \frac{\partial \tilde{\Phi}}{\partial \boldsymbol{\tilde{f}}}  \nonumber  \\ 
           \boldsymbol{B}^T \boldsymbol{L} \boldsymbol{B} \boldsymbol{z}  &= - \frac{1}{2} 
        \boldsymbol{B}^T \boldsymbol{K}^T \boldsymbol{G}
        \frac{\partial \tilde{\Phi}}{\partial \boldsymbol{\tilde{f}}}. \nonumber 
\end{align}
Above, we made use of the linear relationships $\boldsymbol f = \boldsymbol{Kx} $, $\boldsymbol x = \boldsymbol{Bz} $, and $\boldsymbol{\tilde{f}} = \boldsymbol{G}_9 \boldsymbol{f} $ to derive the chained partial derivatives   $\frac{\partial \boldsymbol{x}}{\partial \boldsymbol{z}}, \frac{\partial \boldsymbol{f}}{\partial \boldsymbol{x}},   \frac{\partial \boldsymbol{\tilde{f}}}{\partial \boldsymbol{f}} $. The last unknown $  \frac{\partial \tilde{\Phi}}{\partial \boldsymbol{\tilde{f}}}$ changes each local-global iteration and can efficiently be computed during the local step when looping through each cluster:
\begin{align}
    \frac{\partial{\tilde{\Phi}}}{\partial \boldsymbol{\tilde{f}}} =\text{vec}(
    \begin{bmatrix} 
     \vdots \\
     - \frac{\partial}{\partial \boldsymbol{\tilde{F}}_c} m_{c} \mu_{c} tr(  \boldsymbol{\tilde{F}}^T_{c}\boldsymbol{\tilde{R}}_{c}) \\
     \vdots
    \end{bmatrix}) =
    \text{vec}(
    \begin{bmatrix} 
  \vdots \\  - m_{c} \mu_{c} \boldsymbol{R}_{c} \\ \vdots
    \end{bmatrix})  \nonumber  
    \\
     \text{vec}\left( 
    \ \begin{bmatrix} 
  \vdots \\   -m_{c} \mu_{c} \boldsymbol{R}_{c} \\ \vdots
    \end{bmatrix}\ \right)
\end{align}

\subsection{Hyper-Reduced Co-Rotational Elasticity}
\label{sec:appendix-hyper-reduced-corotational-elasticity}
A discrete linear co-rotational elastic energy is defined as
\begin{align}
    E_{CoRot}(\boldsymbol{u}) = \sum_t^{k} m_t  \mu_t || \boldsymbol{F}_t -  \boldsymbol{R}_t||^2_F +  m_t \frac{\lambda_t}{2}tr^2(\boldsymbol{R}_t^T \boldsymbol{F}_t - \boldsymbol{I}) \nonumber 
\end{align}
Where $\boldsymbol{F}_t$ is our per-tet deformation gradient (a linear function of our degrees of freedom) and $\boldsymbol{R}_t$ is a best-fit rotation on the deformation gradient, a non-linear function of our degrees of freedom.
Expanding the square Frobemius norm, we can identify a quadratic component and non-linear component to our energy:
\begin{align}
    & E_{CoRot}(\boldsymbol{u}) = \nonumber  \\
    & \underbrace{\sum_t^{k} m_t \mu_t tr(\boldsymbol{F}_t^T\boldsymbol{F}_t)}_{\Psi}  + \underbrace{ \sum_t^{k}- 2 m_t \mu_t tr(\boldsymbol{F}^T_t \boldsymbol{R}_t)  \nonumber +   m_t \frac{\lambda_t}{2}tr^2(\boldsymbol{R}_t^T \boldsymbol{F}_t - \boldsymbol{I})}_{\Phi} \\
     &=  \Psi(\boldsymbol{u})  + \Phi(\boldsymbol{u}).  \nonumber 
\end{align}
We can rewrite this energy in its hyper-reduced form by recalling $\boldsymbol{x} = \boldsymbol{Bz}$ and approximating $\Phi \approx \tilde{\Phi}$. We expand our clustered non-linear component to the elastic energy:
\begin{align}
    E_{CoRot}(\boldsymbol{z}) &=   \Psi(\boldsymbol{z})  + \underbrace{\sum_c^{\mathcal{C}}   - 2 m_c \mu_c tr(\boldsymbol{F}^T_c \boldsymbol{R}_c)  \nonumber +   m_c \frac{\lambda_c}{2}tr^2(\boldsymbol{R}_c^T \boldsymbol{F}_c - \boldsymbol{I})}_{\tilde{\Phi}} \nonumber \\ 
     &=  \Psi(\boldsymbol{z})  + \tilde{\Phi}(\boldsymbol{z}).  \nonumber
\end{align}

We then expose the source of the non-linearities, the best fit per-cluster rotation matrices, as auxillary degrees of freedom.
\begin{align}
    E_{CoRot}(\boldsymbol{z}, \boldsymbol{\tilde{R}}) &=    \Psi(\boldsymbol{z})  + \tilde{\Phi}(\boldsymbol{z}, \boldsymbol{\tilde{R}} ).  \nonumber 
\end{align}
The hyper reduced local step becomes:
\begin{align}
    \boldsymbol{\tilde{R}} &=  \argmin_{\boldsymbol{\tilde{R}}}   \tilde{\Phi}(\boldsymbol{z}, \boldsymbol{\tilde{R}} ) \nonumber \\ 
    &= \argmin_{\boldsymbol{\tilde{R}}} \sum_c^{\mathcal{C}} m_c  \mu_c tr(  \boldsymbol{F}^T_c \boldsymbol{R}_c) +  m_c \frac{\lambda_c}{2}tr^2(\boldsymbol{R}_c^T \boldsymbol{F}_c - \boldsymbol{I}) \nonumber \\
    &= \text{PolarSVD}(\boldsymbol{\tilde{F}})
    \nonumber 
\end{align}
which is found through polar decomposition of the per-cluster deformation gradient $\boldsymbol{\tilde{F}}$, obtainable via a reshape operation of $\boldsymbol{\tilde{f}} = \boldsymbol{x}^T \boldsymbol{K}^T \boldsymbol{G} $ .

The hyper reduced global step minimizes:
\begin{align}
      \boldsymbol{z} &= \argmin_{\boldsymbol{z}} \Psi(\boldsymbol{z})   + \tilde{\Phi}(\boldsymbol{z}, \boldsymbol{\tilde{R}} ) \nonumber \\
       &= \argmin_{\boldsymbol{z}} \boldsymbol{z}^T \boldsymbol{B}^T \boldsymbol{K}^T \mathcal{U} \boldsymbol{V} \boldsymbol{K} \boldsymbol{B} \boldsymbol{z} +  \tilde{\Phi}(\boldsymbol{z}, \boldsymbol{\tilde{R}} ) \nonumber \\
         &= \argmin_{\boldsymbol{z}} \boldsymbol{z}^T \boldsymbol{B}^T \boldsymbol{L} \boldsymbol{B} \boldsymbol{z} + \tilde{\Phi}(\boldsymbol{z}, \boldsymbol{\tilde{R}} ) \nonumber 
\end{align}
where we introduce the heterogeneous Laplacian matrix $\boldsymbol{L} = \boldsymbol{K}^T \mathcal{U} \boldsymbol{V} \boldsymbol{K}$ for brevity. Here, $\boldsymbol{K} \in \mathbb{R}^{9k \times 3n} $ is the vector gradient operator over our mesh, mapping deformed coordinates to per-tet deformation gradients, $\boldsymbol{V} \in \mathbb{R}^{9k \times 9k}$ is a diagonal matrix of tetrahedron volumes, $\mathcal{U}  = (\boldsymbol{I}_9  \otimes \text{diag}(\boldsymbol{\mu})) \in \mathbb{R}^{9k \times 9k}$ is a diagonal containing first lamé parameter $\mu_t$ for each tet $t$ along its diagonal. 

The above optimization can be performed without recomputing full space second order terms with an iterative Quasi-Newton method, where each timestep we solve for the search direction $\boldsymbol{z}_{j+1}  = \boldsymbol{z}_{j} + \alpha d\boldsymbol{z} $:
\begin{align}
        \boldsymbol{0} &=  2 \boldsymbol{B}^T \boldsymbol{L} \boldsymbol{B} d\boldsymbol{z} + 2 \boldsymbol{B}^T\boldsymbol{L} \boldsymbol{B} \boldsymbol{z}_j + 
        \frac{\partial \boldsymbol{f}}{\partial \boldsymbol{z}}
        \frac{\partial \boldsymbol{\tilde{f}}}{\partial \boldsymbol{f}} 
        \frac{\partial \tilde{\Phi}}{\partial \boldsymbol{\tilde{f}}} \nonumber  \\
        &=    \boldsymbol{B}^T \boldsymbol{L} \boldsymbol{B} d\boldsymbol{z} + 
         \boldsymbol{B}^T \boldsymbol{L} \boldsymbol{B} \boldsymbol{z}_j +
        \frac{1}{2} \boldsymbol{B}^T \boldsymbol{K}^T \boldsymbol{G}
        \frac{\partial \tilde{\Phi}}{\partial \boldsymbol{\tilde{f}}}  \nonumber  \\ 
           \boldsymbol{B}^T \boldsymbol{L} \boldsymbol{B} d\boldsymbol{z}  &= -  \boldsymbol{B}^T \boldsymbol{L} \boldsymbol{B} \boldsymbol{z}_j - \frac{1}{2} 
        \boldsymbol{B}^T \boldsymbol{K}^T \boldsymbol{G}
        \frac{\partial \tilde{\Phi}}{\partial \boldsymbol{\tilde{f}}}\nonumber 
\end{align}
where we have ignored any of the second order terms buried $\frac{\partial \tilde{\Phi}}{\partial \boldsymbol{\tilde{f}}} $, which would normally incur a full space update in the system matrix every iteration. Because the search direction is no longer \emph{perfect}, we make use of a back-tracking line search to figure out the step size we take along that search direction $\boldsymbol{z}_{j+1}  = \boldsymbol{z}_{j} + \alpha d\boldsymbol{z} $.

Above, we made use of the linear relationships $\boldsymbol f = \boldsymbol{Kx} $, $\boldsymbol x = \boldsymbol{Bz} $, and $\boldsymbol{\tilde{f}} = \boldsymbol{G}_9 \boldsymbol{f} $ to derive the chained partial derivatives   $\frac{\partial \boldsymbol{x}}{\partial \boldsymbol{z}}, \frac{\partial \boldsymbol{f}}{\partial \boldsymbol{x}},   \frac{\partial \boldsymbol{\tilde{f}}}{\partial \boldsymbol{f}} $. The last unknown $  \frac{\partial \tilde{\Phi}}{\partial \boldsymbol{\tilde{f}}}$ changes each local-global iteration and can efficiently be computed during the local step when looping through each cluster:

\begin{align}
    \frac{\partial{\tilde{\Phi}}}{\partial \boldsymbol{\tilde{f}}} &=\text{vec}\left( 
    \begin{bmatrix} 
     \vdots \\
     \frac{\partial}{\partial \boldsymbol{\tilde{F}}_c} -m_{c} \mu_{c} tr(  \boldsymbol{\tilde{F}}^T_{c}\boldsymbol{\tilde{R}}_{c}) +  m_c \frac{\lambda_c}{2}tr^2(\boldsymbol{R}_c^T \boldsymbol{F}_c - \boldsymbol{I}) \nonumber  \\
     \vdots
    \end{bmatrix}\right) \nonumber \\ &=
    \text{vec}\left( 
    \ \begin{bmatrix} 
  \vdots \\   -m_{c} \mu_{c} \boldsymbol{R}_{c} + m_{c} \frac{\lambda_c}{2} \boldsymbol{R}_c tr(\boldsymbol{R}_c^T \boldsymbol{F}_c - \boldsymbol{I}) \\ \vdots
    \end{bmatrix}\ \right) 
\end{align}

\subsection{Adding Hyper-Reduced Inertia}
To add dynamics to our simulation, we add a Kinetic energy term to the elastic energies mentioned in the previous section. We employ a Backward Euler \cite{BaraffWitkin, Liu:2013:FastMassSprings} time integrator:
\begin{align}
E_{Kinetic} = \frac{1}{2h^2}(\boldsymbol{x} -  \boldsymbol{y})^T \boldsymbol{M}(\boldsymbol{x} -  \boldsymbol{y}) \nonumber 
\end{align}
Where $\boldsymbol{y}$ represents position history $\boldsymbol{x}_{hist} = 2\boldsymbol{x}_{curr} - \boldsymbol{x}_{prev}$ and $h$ is the length of the timestep chosen.

We show we can represent this energy fully in our reduced space, without having to go to the full space. 
First we directly rewrite the final displacement in terms of our reduced complementary displacement, and our rigged possition.
$\boldsymbol{x} = \boldsymbol{Bz} + \boldsymbol{Jp}$

\begin{align}
E_{Kinetic}(\boldsymbol{z}, \boldsymbol{p}) &= \frac{1}{2h^2}( \boldsymbol{Bz} + \boldsymbol{Jp}  -  \boldsymbol{x}_{hist})^T \boldsymbol{M}( \boldsymbol{Bz} + \boldsymbol{Jp}  -  \boldsymbol{x}_{hist})  \nonumber 
\end{align}
Expanding this out and only keeping the terms that depend on our degrees of freedom $\boldsymbol{z}$ leads to:
\begin{align}
 &= \frac{1}{2h^2}\boldsymbol{z}^T \boldsymbol{B}^T\boldsymbol{MB}\boldsymbol{z}  +  \frac{1}{h^2}(\boldsymbol{z}^T \boldsymbol{B}^T \boldsymbol{M} \boldsymbol{J} \boldsymbol{p}  - \boldsymbol{z}^T \boldsymbol{B}^T \boldsymbol{M}  \boldsymbol{x}_{hist}) 
\end{align}
Writing $\boldsymbol{x}_{hist}$ in terms of our subspace and control rig: 
\begin{align}
\boldsymbol{y} = \boldsymbol{B} \boldsymbol{z}_{hist} + \boldsymbol{J} \boldsymbol{p}_{hist} \nonumber
\end{align}

Where $\boldsymbol{z}_{hist} = 2\boldsymbol{z}_{curr} - \boldsymbol{z}_{prev}$ and $\boldsymbol{p}_{hist} = 2\boldsymbol{p}_{curr} - \boldsymbol{p}_{prev}$ are our reduced space complementary and rig displacement histories respectively.

Plugging the above into our reduced energy so far:
\begin{align}
 &= \frac{1}{2h^2}\boldsymbol{z}^T \boldsymbol{B}^T\boldsymbol{MB}\boldsymbol{z}  +  \frac{1}{h^2} \boldsymbol{z}^T \boldsymbol{B}^T \boldsymbol{M} \boldsymbol{J}  (\boldsymbol{p} - \boldsymbol{p}_{hist}) -  \frac{1}{h^2} \boldsymbol{z}^T \boldsymbol{B}^T \boldsymbol{M} \boldsymbol{B} \boldsymbol{z}_{hist}
\end{align}

Adding this kinetic energy to the prior elastic energies, then minimizing would simply mean adding $  \frac{1}{h^2}\boldsymbol{B}^T \boldsymbol{M} \boldsymbol{B}$ to the quadratic terms, and then 
$ \boldsymbol{f}_{inertia} =  \frac{1}{h^2}\boldsymbol{z}^T \boldsymbol{B}^T \boldsymbol{M} \boldsymbol{J}  (\boldsymbol{p} - \boldsymbol{p}_{hist}) - \frac{1}{h^2}\boldsymbol{z}^T \boldsymbol{B}^T \boldsymbol{M} \boldsymbol{B} \boldsymbol{z}_{hist}$
to the linear terms, as shown in  Algorithm \ref{alg:simulationStep}.

\end{document}